\documentclass[final,3p,times,authoryear]{elsarticle}

\usepackage{amssymb}
\usepackage{hyperref}
\hypersetup{colorlinks=true,citecolor=blue}
\usepackage{graphicx,amsmath}
\usepackage{ifthen}
\usepackage{multirow}
\usepackage{breakurl}
\def\beq{\begin{equation}}
\def\eeq{\end{equation}}
\def\bea{\begin{eqnarray}}
\def\eea{\end{eqnarray}}

\biboptions{round,comma,sort&compress}

\journal{Environmental Innovation and Societal Transitions}

\begin{document}
\begin{frontmatter}

\title{The behavioural aspect of green technology investments: a general positive model in the context of heterogeneous agents}

\author[CG]{F. Knobloch  }
\ead{knobloch@cantab.net}
\author[RU,CG]{J.-F. Mercure \corref{cor1}}

\address[CG]{Cambridge Centre for Environment, Energy and Natural Resource Governance (C-EENRG), University of Cambridge, 19 Silver Street, Cambridge, CB3 1EP, United Kingdom}

\address[RU]{Department of Environmental Science, Radboud University, PO Box 9010, 6500 GL Nijmegen, The Netherlands}

\cortext[cor1]{Corresponding author: Florian Knobloch}

\begin{abstract}

Studies report that firms do not invest in cost-effective green technologies. While economic barriers can explain parts of the gap, behavioural aspects cause further under-valuation. This could be partly due to systematic deviations of decision-making agents' perceptions from normative benchmarks, and partly due to their diversity. This paper combines available behavioural knowledge into a simple model of technology adoption. Firms are modelled as heterogeneous agents with different behavioural responses. To quantify the gap, the model simulates their investment decisions from different theoretical perspectives. While relevant parameters are uncertain at the micro-level, using distributed agent perspectives provides a realistic representation of the macro adoption rate. The model is calibrated using audit data for proposed investments in energy efficient electric motors. The inclusion of behavioural factors reduces significantly expected adoption rates: from 81\% using a normative optimisation perspective, down to 20\% using a behavioural perspective. The effectiveness of various policies is tested.

\end{abstract} 


\end{frontmatter}


\section{Introduction}

Why are firms' investments in green technologies lower than predicted by engineering studies? Can the gap be sufficiently explained as a rational response to risky future cost-savings and information asymmetries? Or do behavioural aspects cause a further systematic undervaluation of green technologies compared to contemporary mainstream investment theory? If so, what implication does this have for policies aimed at increasing such investments?

A green technology is one that \emph{generates or facilitates a reduction in environmental externalities relative to the incumbent} \citep[p.~2]{Allan:2014}. When this is achieved by reducing the input for a given output, a technology potentially reduces operating costs along with the externality. According to mainstream economic theory, a profit-maximizing firm should undertake such an investment whenever these future savings outweigh the upfront cost. 

However, studies based on engineering data regularly report that seemingly cost-effective green investments are not undertaken \citep[for a discussion,  see][p. 132]{Grubb:2014}. A reduction of environmental externalities at apparently negative cost is consistently found within the literature  \citep{Allan:2014}.  \cite{McKinsey:2009} claims that global CO$_{2}$ emissions could be reduced by 11Gt per year by investing in cost-effective green technologies, which is not currently happening. The largest potential is attributed to energy efficiency measures, and the gap between realised market outcomes and the normative cost-minimising benchmark is referred to as the \emph{energy efficiency gap}  \citep{Jaffe:1994}.  

Considering the gradual process of technology diffusion \citep[e.g.][]{Geels:2002, Rogers:2010}, it is meaningful to analyse how firms decide on a green technology investment when it is directly presented to them, such as after an energy audit. Firms apparently reject a large fraction of seemingly cost-effective project recommendations, as seen from external engineering perspectives \citep{Anderson:2004}. Firms dismissing profitable investments in such a systematic way makes effective policy-making for stimulating technological change difficult: it becomes unclear how to create fruitful incentives for technology adoption. In the case of technologies that reduce energy use and emissions, this has important implications for climate change mitigation policy  \citep{Worrell:2009}.

There exists a vast quantity of literature that both analyses the scope of the gap, and suggests possible reasons --- the so-called \emph{market barriers}  \citep{Sorrell:2004, Sorrell:2011}. Many studies attempt to realign the observed adoption gap with the neoclassical theory paradigm, in which the representative agent adopts cost-minimizing or utility-maximizing measures. For example,  \cite{Sutherland:1991} argues that firms have rational reasons to reject green technologies, but that these reasons are mostly omitted in engineering studies --- such as hidden costs, risk, imperfect information and capital constraints. As a result, many investments may be less profitable than they seem to be.\footnote{The issue with such a line of reasoning is that it allows accepting the use of an un-falsifiable theory, where cause is attributed to unknown variables.}

While economic barriers can explain parts of the gap, others question the behavioural realism of decision mechanisms assumed in theories \citep[e.g.][]{DeCanio:1998, Gillingham:2014}. First, firms might not act as profit-maximizers, but instead look for satisfactory solutions \citep{Simon:1955}. Second, behavioural economics shows that human decisions systematically violate the axioms of expected utility theory, and are better described based on psychological foundations \citep{Kahneman:1979, Tversky:1992}. Since decisions of firms are a combination of human decisions, they can be subject to the same behavioural biases \citep{Grubb:2014}. 

The relevance of behavioural factors for climate and energy policy is now widely acknowledged \citep{Pollitt:2012, Allan:2014}. However, there is limited knowledge on what drives people's and firms' behaviour, and how this influences aggregate outcomes. 

This is particularly relevant in the perspective of sustainability transitions studies, in which the process of decision-making by agents is often not emphasised, but is at least as important as cultural, regulatory and other contextual factors that influence or limit the formation of new socio-technical regimes \citep[e.g. as in][]{Geels:2002}. This type of research can improve representations of agent behaviour, quantitatively and qualitatively, in the various representations of the field (i.e. the multi-level perspective and technology innovation systems), since it is ultimately agent adoption choices that determine the successful replacement of old socio-technical regimes by new ones, and the diffusion of innovations out of their niches.

To clarify current understanding and to improve quantitative analysis, we combine known behavioural facts into a simple aggregate model of decision-making by heterogeneous agents for technology adoption. This allows identification and quantification of relevant barriers and behavioural factors, without recourse to unknown variables, and aims at a higher predictive power when modelling the adoption of a green technology. Finally, it can provide key insight on the likely effectiveness of policies from a behavioural perspective. 

The diverse perception by agents of the profitability of a technology can influence rates of uptake. Diversity implies varying adoption thresholds across agents, and is thus partly responsible for the typically observed gradual diffusion of innovations  \citep{Rogers:2010}. Thus, within this model, investment decisions are simulated based on the technological and behavioural diversity of firms. It is assumed that a technology has different costs and benefits, as perceived by every single firm. Due to heterogeneous decision-making parameters, investment decisions by firms differ even when faced with the same problem and data. Furthermore, behavioural aspects include systematic biases in firms' perceptions of technological opportunities. This is typically interpreted with prospect theory \citep{Kahneman:1979}, in which gains and losses are not valued equally. Thus we combine here both effects of \emph{behavioural diversity} and of \emph{behavioural biases}. 

To quantify the relevance of behavioural factors compared to a normative benchmark, the model simulates technology adoption using three different possible types of decision-making --- referred to as \emph{levels of decision-making}: optimizing, satisficing and behavioural. Each level corresponds to a different point of view on the investment decision, or method for project evaluation, and thus includes different barriers and degrees of heterogeneity.  The model can be used to estimate and compare the effectiveness of policies according to different levels of decision-making, or theoretical paradigms from the viewpoint of the modeller.

To demonstrate the model's abilities, we apply it to a case study of electric motors, which account for 43--46\% of global electricity consumption. It is estimated that cost-effective investments could increase their average efficiency by 20--30\% \citep{IEA:2011}. Nevertheless, firms systematically reject these investment possibilities (\emph{ibid}). To explain this phenomenon, the model simulates these investment decisions on different levels of decison-making. The results are compared to observed decisions that have been taken by firms after energy audits. 

Policies can be applied to influence the rate of uptake of green technologies, and we suggest that these can take two forms. In the traditional sense, financial incentives such as taxes and subsidies can be applied to improve the rate of uptake --- a rate that depends on the level at which decision-making is made, i.e. the method of project assessment by firms in a particular situation. However, policies can also seek to change the way in which project assessment is carried out, \emph{shifting the level at which decisions are taken}, using methods such as information campaigns. We show that for electric motors, this could have a high impact on rates of uptake. 

\section{Literature Review and Background}

The field of sustainability transitions studies (STS) provides a powerful, if mostly qualitative, rationale to describe and connect the components influencing the diffusion of green innovations out of their existing niches to the mainstream \citep[e.g.][]{Geels:2002,Geels2005,Genus2008,Geels2011} and how to foster transitions \citep[e.g.][]{Rotmans2001}. Little quantitative work, however, characterises how these processes collectively act to make a transition happen \citep[see e.g.][]{Holtz2011}. A field of research is emerging for modelling sustainability transitions \citep[for a discussion, see][]{Holtz2015}. There, emphasis is given to path-dependence, complexity and interactions across domains, but perhaps insufficient attention is given to characterise decision-making at the level of agents. Meanwhile, the parent field of evolutionary economics also considers issues of technological change; and there, more emphasis has been given to quantitative methods \citep[e.g. see ][]{Safarzynska2012, Mercure2014, Saviotti1995}, even if no dominant model yet exists. But concepts going beyond standard utility theory have generally been considered at the core of evolutionary theories and models of technological change, in particular bounded rationality \citep{Metcalfe1988, Saviotti1991}, enabling quantitative considerations to be made. This work thus positions itself within an evolutionary economics perspective; but many of these evolutionary elements could also be considered in STS \citep[e.g. see the comparison by ][]{Safarzynska2012b}.

Technology diffusion describes the gradual adoption of innovations by firms and consumers, which involves choices to be made. Adopters first have to learn about the new technology's existence and benefits, and then decide whether to adopt or not. Considering heterogeneous agents in this context implies a gradual diffusion process, as different agents have different thresholds to their decisions \citep{Rogers:2010}. Many studies show that the process regularly resembles a path similar to logistic curves \cite[e.g.][]{Grubler:1991, Mansfield1961, Fisher1971, Marchetti1978, Grubler1999, Nakicenovic1986}. \cite{Wilson:2011} identify seven `\emph{grand patterns}' of technological transitions. \cite{Geels:2002} develops the multi-level perspective to understand technological transitions in a complex societal context, in which components of incumbent socio-technical regimes (e.g. infrastructure, knowledge, regulatory context, culture and user practices) influence their successful or unsuccessful diffusion. However, in all of these analysis perspectives, the microeconomic picture remains to some degree unclear.

\begin{figure}[t]
	\begin{center}
		\includegraphics[width=.7\columnwidth]{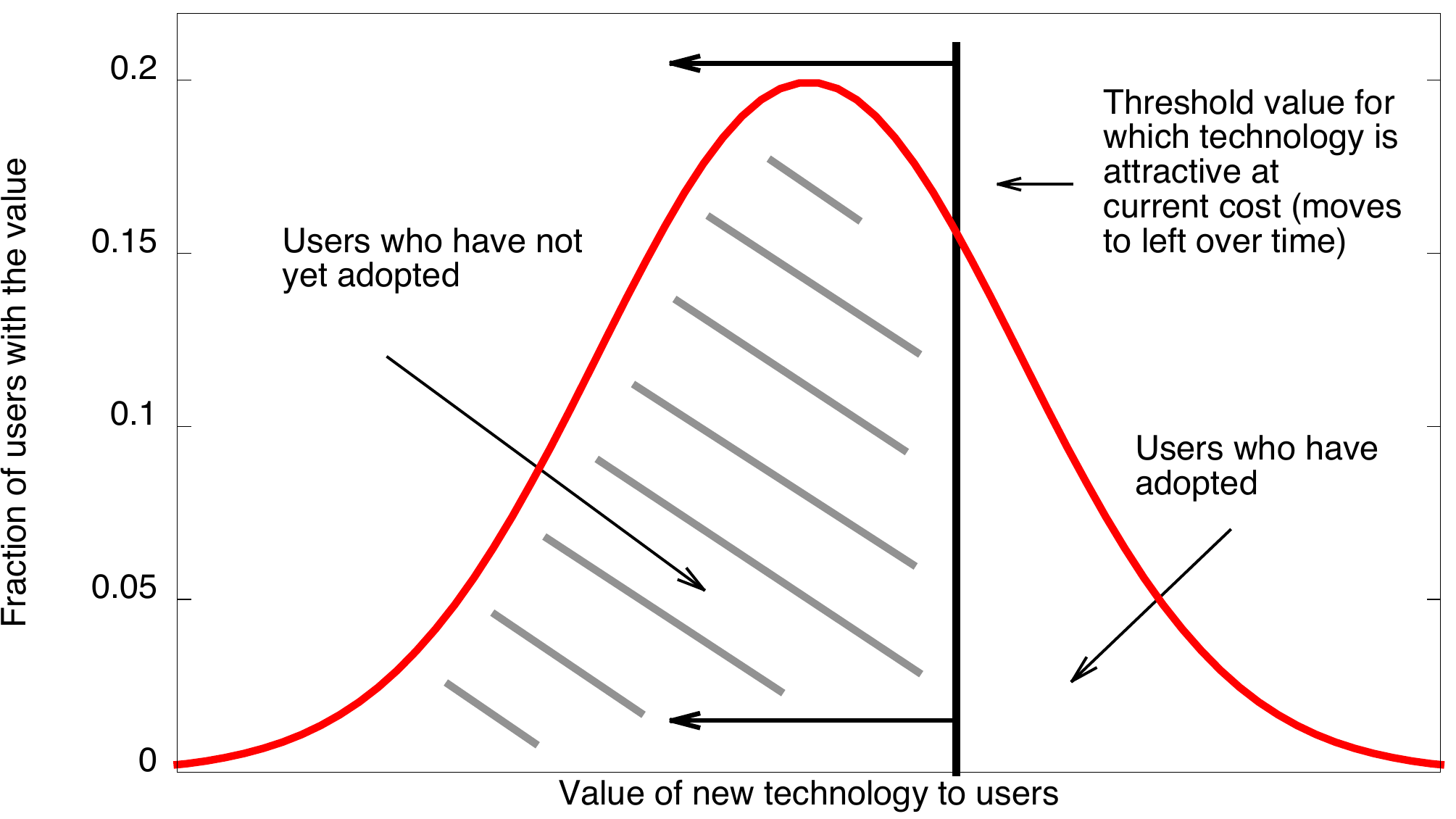}
	\end{center}
	\caption{Diffusion with heterogeneous adopters. Adapted from \cite{Allan:2014}.}
	\label{fig:Figure1}
\end{figure}

From a microeconomic perspective, a gradual diffusion can be related to changing costs or benefits as perceived by heterogeneous adopters with different perspectives (see figure~\ref{fig:Figure1}). Various processes may change perceived costs and benefits over time, such that the adoption becomes increasingly attractive. As emphasised by \cite{Rogers:2010}, heterogeneous characteristics of agents (e.g. risk-aversion, tastes and income) imply different net benefits of adopting a new technology. With learning-by-doing, gradual cost reductions occurring with cumulative adoptions lead to a self-reinforcing phenomenon \citep{Arthur:1989}.

Not adopting a green technology deemed profitable, however, constitutes a gap between classical theory and reality. It can be explained either by assuming that agents act rationally, but don't invest in seemingly profitable technologies due to economic factors that are omitted in engineering studies --- so called market barriers. Or, agents systematically violate the rationality axioms of expected utility theory --- which would constitute \emph{irrational} behaviour from a neoclassical perspective.

Some studies attempt to explain such recurrent efficiency gaps by evoking purely economic factors. \cite{Sutherland:1991} argues that the rejection of green technology investments represents `\emph{real costs in real markets}': given that many such investments are illiquid and undiversifiable, their rejection is a rational response to risk. The reliance on  `\emph{overlooked}' economic factors is a common explanation in the literature. However, as \cite{Jaffe:1994} argue, ``\emph{such explanations must advance beyond the tautological assertion that (...) there must be some unobserved adoption costs}'' (p. 805), which effectively impedes constructive scientific thought in this context. 

\cite{Sorrell:2004} classify different classes of barriers. \cite{Sorrell:2011} is widely credited with the standard taxonomy, used here as fundamental building blocks for our theory.\footnote{Although the authors focus on the specific case of energy efficiency investments, the classification can easily be applied to any green technology.} Barriers are grouped into six classes (table~\ref{tab:barriers}), and defined as ``\emph{a postulated mechanism that inhibits a decision or behaviour that appears to be both energy efficient and economically efficient}'' (p. 4).  

\begin{table}[t]\footnotesize
\begin{center}
		\begin{tabular*}{1\columnwidth}{@{\extracolsep{\fill}} p{3.0cm}| p{12.5cm} }
			
			Barrier & Summary \\
			\hline
			\hline
            Risk & The short paybacks required for energy efficiency investments may represent a rational response to risk. This could be because energy efficiency investments represent a higher technological or financial risk than other types of investment, or that business and market uncertainty encourages short time horizons. \\
            \hline
            
            Imperfect information  & Lack of information on energy efficiency opportunities may lead to cost-effective opportunities being missed. In some cases, imperfect information may lead to inefficient products driving efficient products out of the market. \\
            \hline
            
             Hidden costs  & Engineering-economic analyses may fail to account for either the reduction in utility associated with energy efficient technologies, or the additional costs associated with them. As a consequence, the studies may overestimate energy efficiency potential. Examples of hidden costs include overhead costs for management, disruptions to production, staff replacement and training, and the costs associated with gathering, analysing and applying information.  \\
            \hline
            
            Access to capital  & If an organisation has insufficient capital through internal funds, and has difficulty raising additional funds through borrowing or share issues, energy efficient investments may be prevented from going ahead. Investment could also be inhibited by internal capital budgeting procedures, investment appraisal rules and the short-term incentives of energy management staff. \\
          
           \hline
             Split incentives  & Energy efficiency opportunities are likely to be foregone if actors cannot appropriate the benefits of the investment. For example, if individual departments within an organisation are not accountable for their energy use they will have no incentive to improve energy efficiency. \\
            \hline
            
            Bounded rationality  & Owing to constraints on time, attention, and the ability to process information, individuals do not make decisions in the manner assumed in economic models. As a consequence, they may neglect opportunities for improving energy efficiency, even when given good information and appropriate incentives.  \\
            \hline
			
		\end{tabular*}
	\caption{Barriers to energy efficiency, following \cite{Sorrell:2011}.}
	\label{tab:barriers}
\end{center}
\end{table}

The central microeconomic theory of choice under risk is expected utility theory (EUT) by \cite{Von-Neumann:1947}. EUT can be seen as either a normative or a positive theory; however seeing it as positive requires accepting heroic assumptions over agents thought processes (rationality). 
Meanwhile, as a normative theory, it is not restricted in this sense, and describes a subjective set of objectives. Nevertheless, economists widely apply EUT as a positive description of human behaviour, arguing their view in this way: so long as deviations by agents from rationality are random and unsystematic, they are irrelevant on average. Empirical evidence shows, however, that many deviations from EUT are systematic\footnote{Prominent examples being the choice paradoxes analysed by \cite{Allais:1953} and \cite{Ellsberg:1961}} as well as correlated, thus not cancelling out without any impact. 

Since most decisions of organisations are made by individuals, firms display comparable behavioural characteristics \citep{Shogren:2008, Grubb:2014}. Consequently, the assumption of profit-maximization may be an oversimplified representation of a firm's investment behaviour. 

\cite{Gillingham:2009} identify three key areas from behavioural economics that are most relevant for energy efficiency policy: bounded rationality, heuristic decision-making and prospect theory. 

\emph{Bounded rationality} \citep{Simon:1955} replaces unbounded maximization with satisficing behaviour. Agents are assumed to behave rationally, but within cognitive limits. Manifestations are \emph{heuristics} in the form of \emph{rules of thumb} or the replacement of an originally complex decision problem with a simpler, roughly accurate one. One possible implication is that managers maximize their own utility within the constraint of achieving \emph{satisficing} (instead of maximal) profits \citep{Williamson:1967}. For investments, this may explain why many firms use simplified payback thresholds as a key decision criterion (instead of net-present-value calculations). Even small individual discrepancies between satisficing and optimizing solutions can result in large deviations of overall market allocations \citep{Akerlof:1985}. 

Bounded rationality is a concept quite central to evolutionary economics and evolutionary game theory. Having originated in evolutionary biology \citep{Hofbauer1998}, evolutionary game theory offers a powerful method to explore social dynamics (e.g. \citealt{Safarzynska2010, Saviotti1995}; for a review, see \citealt{Safarzynska2010}), for instance based on replicator dynamics as a dynamical equation. In that framework, social dynamics are explained partly through bounded rational decision, partly through multi-agent interactions (e.g. social influence). This generally results in complex dynamical models \citep[with attractors; e.g. see the discussion on increasing returns by ][]{Arthur:1989}.

\cite{Kahneman:1979} proposed \emph{prospect theory }as a formal descriptive framework for decision analysis. They systematically mapped observed contradictions to EUT and conclude that for most decisions, optimality considerations are less relevant than the question, ``\emph{What course of action seems most natural in this situation?}'' \citep[p.~1469]{Kahneman:2003}. For example, in empirical studies, losses are weighed roughly twice as much as gains \citep{Greene:2011} --- termed as \emph{loss aversion}. Overall, prospect theory helps to explain various behavioural biases that are potentially relevant for green technology investments. 

Both bounded rationality and prospect theory provide possible explanations for the energy efficiency gap. Both are consistent with empirical observations that the sensitivity to future cost savings is much lower than to upfront costs: \cite{Jaffe:1994} empirically analyse the diffusion of thermal insulation and find investments to be three times more sensitive to upfront subsidies than to the energy price. \cite{Hassett:1995} find that investments in energy conservation are eight times more sensitive to tax credits than to electricity price increases. \cite{Anderson:2004} analyse manufacturing companies' investment decisions after energy audits and report a sensitivity towards upfront payments that is 40\% larger than to annual savings. 

For a perfectly rational investor, a relative change in upfront costs should have the same effect on investment decisions as a change in discounted expected future payoffs of the same relative magnitude \citep{Allan:2014}. The observed overweighting of upfront costs could therefore indicate systematic behavioural deviations from EUT. Unfortunately, to reliably diagnose a deviation from utility maximization, researchers have to know the agents' true subjective expectations about future payoffs. In a real world problem, it is likely that a whole variety of subjective decision-making procedures are found, and it would be impossible to reliably enumerate them when studying, for instance, nation-wide energy policy experiences. However, it is possible that simpler explanations, not based on EUT and not chosen \emph{ad hoc}, can be found, which may provide higher explanatory power than EUT --- as we attempt here.


\section{A general positive model and classification of decision-making}

We introduce a model of technology adoption that explicitly includes the diversity of agents and systematic behavioural deviations from EUT. The overarching goal is the simulation of technology adoption and policy outcomes in the context of different decision-making processes of firms, i.e. how firms assess green investment projects. 

\subsection{Methodology} 

The model is based on three closely interrelated building blocks: perceptions, heterogeneity and risk. 

\begin{itemize}
\item \emph{Perceptions} differ between individuals. The same set of data on an investment may be interpreted in different ways by different firms, and by modellers. All models of decision-making are subjective sets of perspectives. 
\item \emph{Heterogeneity} of economic fundamentals and decision-making procedures imply heterogeneous decisions. The adoption of a homogenous technology may be profitable for the average firm, but not for all firms, due to varying contexts (in contrast to varying perceptions). The relevant decision criteria are heterogeneous: firms may have different planning horizons, and may use entirely different sets of decision criteria (decision protocols).
\item \emph{Risk
} affects firms' decisions. From the firms' perspective, the profitability of an investment is a risky outcome. This risk is not necessarily an objective property of the world, but can be seen as a subjective or psychological conception \citep{Crawford-Brown:1999}. 

\end{itemize}

Given that firms and individuals within them apply heterogeneous decision protocols, it is generally not possible to forecast individual choices. However, possible values of parameters used for decisions across individuals in a group are bounded, and based on plausible assumptions on parameter distributions, it is possible to construct an overview of how the choices are distributed among a sufficiently large number of firms, as follows.

We define three levels of decision-making, plus a benchmark level for a representative firm (see table~\ref{tab:levels}). Each level corresponds to the perspective and decision criteria of a specific hypothetical actor: the engineer, optimizing and satisficing firms (acting according to unbounded and bounded rationality), and individuals within firms. 

We use these definitions to ask the question: if agents were to decide according to the criteria of level~X, what would be the aggregate system level outcome? Different levels will bring different amounts of variations (distributions around EUT) and biases (systematic deviations from EUT). Depending on their framing, particular problems can be assigned to a particular level of analysis, and the model can be used to obtain quantitative insight on likely aggregate outcomes of decision-making. This insight can then be used for two purposes: to estimate the likely effectiveness of particular hypothetical policies, and to classify impacts of various barriers on outcomes.

\begin{table}[h]\footnotesize
\begin{center}
		\begin{tabular*}{1\columnwidth}{@{\extracolsep{\fill}} l l p{3cm} p{8cm}}
			
			Level & Decision behaviour & Actor & Defining question \\
			\hline
			\hline
            0 & Technological & Engineer & What is cost-effective for a representative firm from a technological point of view? \\
            \hline
            1 & Optimizing & Heterogeneous firms with unbounded rationality & What \emph{should be} perceived as individually beneficial for profit-maximizing heterogeneous firms that decide according to the concepts proposed by standard micro-economics? \\
            \hline
            2 & Satisficing & Heterogeneous firms as organisations with bounded rationality & What \emph{should be} perceived as individually beneficial for a profit-maximizing firm, given organisational structures and limited decision resources? \\
            \hline
            3 & Behavioural & Heterogeneous individuals inside heterogeneous organisations & What is \emph{perceived to be} individually beneficial by an agent inside a profit-maximizing firm, given organisational structures, limited cognitive resources and systematic behavioural deviations from EUT? \\
            \hline
			
		\end{tabular*}
	\caption{Levels of decision-making and their analytical relevance.}
	\label{tab:levels}
\end{center}
\end{table}

For each level, a \emph{different subjective decision protocol} is formalised. It is assumed for simplicity that each firm has a binary choice to adopt a green technology or not, which is compared to an alternative \emph{default} technology with respect to perceived investment costs (denoted as $\Delta C$) and perceived future cost-savings (denoted as $\Delta B$). Within the engineering benchmark and the first level, the decision-making protocol involves calculating an investment's \emph{net-present-value} ($NPV$). The decision criteria for levels 2 and 3 are formulated in a similar way, but using an investment's \emph{perceived net-present-benefit} ($NPB$). On a given level, a firm is assumed to invest in the green technology if and only if the  \emph{net-present-value} (on level 0 and 1) or \emph{perceived net-present-benefit} (on levels 2 and 3) is larger than zero.

In section \ref{sect:CaseStudy}, the decision protocols are calibrated using a case study on electric motors in industry. Based on audit data, probability distributions are assigned to all heterogeneous parameters. The outcome variable, \emph{net-present-value/benefit}, is a combination of all heterogeneous parameters. In order to combine the heterogeneity of the superposed behavioural layers, the model evaluates combined distributions of $NPB$ up to each level of decision-making by performing Monte Carlo simulations.\footnote{This can equally be done by calculating convolutions of probability distributions with one another.}

The results are interpreted as the investment decisions of a heterogeneous population of firms. Each element of Monte Carlo simulations corresponds to the decision of one individual firm. Based on these simulations, the share of adopters and non-adopters is compared and visualised for each level of decision-making. The results can be used to identify the robustness of policies for the levels appropriate to the problem of interest.

\subsection{Levels of decision-making} 

In the following, we describe the levels of decision-making including the respective assumptions and decision protocols. $E_{i}(X_{i})$ refers to firm $i$'s subjective expectation value of a given $X$ variable (e.g. price of electricity) for firm~$i$ (see table~\ref{tab:level1}).

\subsubsection*{(0) Technological}

The technological level is a hypothetical baseline. The green technology investment is analysed from an engineering perspective \citep[e.g. as in][]{McKinsey:2009},\footnote{The perspective of the technology developers, and according to them, how the technology is supposed to be used.} assuming a representative firm that decides according to a perfectly rational decision protocol with infinite amounts of perfectly reliable information taken into consideration (i.e. protocol and information known by the modeller). There is neither risk nor uncertainty.

\begin{table}[t]\footnotesize
\begin{center}
	\begin{tabular*}{1\columnwidth}{@{\extracolsep{\fill}} p{3cm}  p{4cm} p{4cm} p{4cm}}			
		Level & Assumptions & Barriers & Parameters \\
		\hline
		\hline
            0 --- Technological & homogenous agents, \newline no risk or uncertainty, \newline perfect information, \newline 			unbounded rationality  & \emph{none} & \emph{hypothetical baseline}\\
		\hline
            1 --- Optimizing & heterogeneous agents, \newline perfect information, \newline unbounded rationality, \newline undiversifiable risk  & + hidden costs \newline + external risk \newline + business risk \newline + restricted credit \newline + imperfect information 
		& discount rates $(r_{i})$,\newline
		expectations $(E_{i})$, \newline 
		 implicit weights $(\gamma _{i})$ \newline 
		heterogeneity in: \newline
                 cost and benefits $(\Delta C_{i}, \Delta B_{i})$, \newline
                 lifetimes $(n_{i})$\newline \\
		\hline
            2 --- Satisficing & heterogeneous agents, \newline undiversifiable risk, \newline firms as organisations, \newline
		decision constraints & + capital budgeting \newline + split incentives \newline + limited time \newline + limited resources  &
		Payback criterion: \newline critical thresholds ($b_{i}$) \\
		\hline
            3 --- Behavioural & heterogeneous agents, \newline undiversifiable risk, \newline firms as organisations, 
            	\newline deviations from EUT & + loss aversion \newline + status quo bias \newline + values  & 
	         Prospect theory loss aversion function: \newline weighting factors $(\lambda)$, \newline decreasing marginal utility $(\alpha, \beta)$\\     
		\hline
		\end{tabular*}
	\caption{Levels of decision-making and their parameters.}
	\label{tab:level1}
\end{center}
\end{table}

According to neoclassical theory, the appropriate decision protocol for a profit-maximizing firm is a net-present-value (NPV) calculation \citep{Damodaran:2007}: the investment's value is defined as the present value of its expected cash flows. If the NPV is positive, an investment should be undertaken. 

For the case of green technology investments, it is assumed that the relevant cash flows are the upfront investment costs $(\Delta C)$ in the present period $(t=0)$ and annual benefits $(\Delta B)$ throughout the investment's lifetime $(n)$ (from $t=0$ up to $t=n$). Investment costs are either the net cost-difference relative to the default investment (e.g. if retired equipment has to be replaced) or the green technology's gross investment costs (e.g. if the investments replaces current equipment prematurely). Benefits are defined as input cost-savings relative to the default technology: the change in the required quantity $(\Delta q)$ of an input (e.g. electricity) given a constant output, multiplied by the input's market price $(p)$ (inclusive of any fees and taxes). From the technological (the technology producer's) perspective, the green investment is assumed to be risk-free. To account for the time-value and opportunity cost of money, future cash flows are discounted by a discount rate~$r$ (on this level, the rate of return on risk-free investments).  

Given all assumptions, the decision criterion of the representative firm can be described by formula (1):
\beq
NPV := -\Delta C + \sum_{t=0}^{n} \frac{\Delta B_{t}}{(1+r)^{t}} = -\Delta C + \sum_{t=0}^{n} \frac{p_{t}*\Delta q_{t}}{(1+r)^{t}}
\label{sect:eq1}
\eeq
The resulting decision is a point estimate for a representative firm, which is used to predict the homogeneous behaviour of all firms. This is what might be used for instance in cost-optimization models, based on averaged technology costs and benefits.

\subsubsection*{(1) Optimizing}

The optimizing level of decision-making analyses the optimal decisions of firms according to the principles that are suggested by neoclassical theory: how should fully rational firms behave in order to maximize their profits? Compared to the technological level, there are three main differences.

First, heterogeneity is introduced. A technology is assumed to have different upfront costs $(\Delta C_{i})$ and benefits $(\Delta B_{i})$ for different firms. The upfront costs depend on various heterogeneous factors, such as staff costs or disruptions in the production process \citep{Sorrell:2011}. In terms of barriers, these differences are referred to as `\emph{hidden costs}'. Individual benefits differ in a similar way --- both due to differences in input prices $(p_{i})$ and quantities saved $(\Delta q_{i})$ (e.g. due to varying load factors). Furthermore, overall benefits depend on the investment's expected useful lifetime $(E_{i}(n_{i}))$, which may differ between firms due to heterogeneous expectations about the technological lifetime, as well as heterogeneous planning horizons for the use of the technology (e.g. due to a planned factory shut-down). 

Second, firms are assumed to differ with regard to capital access. Future cash flows are therefore discounted by individual firms' private discount rates $(r_{i})$, here defined as a their \emph{weighted average cost of capital }(WACC). In terms of barriers, $r_{i}$ represents `\emph{restricted access to credit}'  (e.g. due to being a high risk borrower). 

Third, risk and imperfect information are introduced. While firm $i$ may know some determinants of future benefits (like its individual load factor), it has to rely on subjective expectation values $(E_{i})$ for others (like the technological lifetime and future input prices). Furthermore, firms may be subject to imperfect or asymmetric information on the technology's true performance: if buyers can't reliably observe promised benefits and sellers can't credibly communicate them, they might be rationally ignored in purchase decisions, resulting in adverse selection \citep{Akerlof:1970}. The credibility of promised cost-savings is likely to be higher if a technology's advantages are well proven or certified, and lower if not directly observable (e.g. due to a lack of metering). 

From the firm's perspective, it could therefore be rational to perform a risk-adjustment of expected benefits \citep{Sutherland:1991} --- depending on individual degrees of risk-aversion, as well as heterogeneous perceptions of the investment's riskiness and credibility. Since all this is highly subjective, we restrain from explicitly modelling the corresponding risk-adjustment by firms. Instead, based on \cite{Allcott:2012}, we define the parameter $\gamma _{i}$ (between 0 and 1) as an implicit weight on expected savings in the agent's decision. From an optimizing point of view, it can be evaluated as a statistic for all \emph{investment inefficiencies}. Since it is hardly possible to calibrate $\gamma _{i}$ beyond an arbitrary level, it is not used in any of the following simulations (so by default set to 1). However, we implicitly estimate $\gamma _{i}$ in the sensitivity analysis by finding which average value of $\gamma _{i}$ is consistent with observed investment behaviour when assuming that firms decide based on level 1.

The resulting decision criterion for heterogeneous optimizing firms is:
\beq
NPV_{i} := -\Delta C_{i} + \sum_{t=0}^{E_{i}(n_{i})} \frac{\gamma _{i} * E_{i}(\Delta B_{i, t})}{(1+r)^{t}} = -\Delta C_{i} + \sum_{t=0}^{E_{i}(n_{i})} \frac{\gamma _{i} * E_{i}(p_{i,t})*E_{i}(\Delta q_{i,t})}{(1+r)^{t}},
\eeq

The defining differences to eq.~\ref{sect:eq1} are to use \emph{subjective agent expectations of the future} $E_{i}(:)$ for random variables, $i$-subscripts for heterogeneity, private discount rates $(r_{i})$ and the implicit parameter $\gamma _{i}$.  The respective parameters are assumed to be different for each firm. This implies heterogeneous investment decisions, which will result in a distribution of net-present-values $(NPV_{i})$. From an optimizing perspective, every firm with $NPV_{i} > 0$ should undertake the investment.

\subsubsection*{(2) Satisficing}

The satisficing level of decision-making focuses on the behaviour of firms as organisations, taking into account organisational constraints and limited resources for decision-making (time and attention). \cite{Graham:2001} survey the budgeting behaviour of firms in the US and conclude that most don't perform NPV calculations. Instead, 57\% `\emph{always or almost always}' use a payback-time criterion, which does neither include discounting, nor risk-adjustments. Most importantly, it ignores all future cash flows after an arbitrary cut-off date. 

To account for this observed behaviour, the decision criterion is replaced with a simple payback criterion. Firms are assumed to compare a green technology's investment costs with its expected future cost-savings for a limited number of years, denoted by individual payback thresholds $b_{i}$. This threshold is commonly found to be between one and five years, with a majority of firms requiring an investment to pay for itself within one to two years \citep{Anderson:2004}. As a result, any future cost-savings beyond the cut-off date are ignored.

Given firms' constraints on decision-making, the replacement of a NPV calculation with a simpler criterion can still be interpreted as being optimal within the concept of bounded rationality. For instance, senior staff may lack the capacity to monitor all projects. When a green technology investment is perceived as relatively unimportant, it may be evaluated under a less demanding basis of payback criteria \citep{Sorrell:2011}. Furthermore, short payback thresholds might be a consequence of split incentives and `managerial risk-aversion' inside an organisation, for example, when the responsible staff is only evaluated based on short-term goals. 

The resulting decision criterion is summarised by eq.~\ref{eq:Org}. 
\beq
NPB_{i} := -\Delta C_{i} + \sum_{t=0}^{b_{i}} E_{i}(\Delta B_{i, t}) = -\Delta C_{i} + \sum_{t=0}^{b_{i}} \left[ E_{i}(p_{i, t}) * E_{i}(\Delta q_{i, t}) \right]
\label{eq:Org}
\eeq
Note that the decision criterion is no longer $NPV_{i}$. To allow for a comparison between levels, it is instead defined as the \emph{net-present-benefit} $(NPB_{i})$ as perceived by individual organisations.

\subsubsection*{(3) Behavioural}

The behavioural level of decision-making adopts a purely positive perspective. In no way does it imply how a decision should be made. Instead, it focuses on the perspective of individuals, working inside heterogeneous organisations --- how \emph{do} individual decision-makers behave? --- taking into account not just organisational structures, but also systematic behavioural deviations from EUT. The decision is seen here from the perspective of the individual who is ultimately responsible for the investment decision, according to internal structures and hierarchies.

According to EUT, a decrease in upfront costs should result in the same utility change as an increase in future benefits of the same relative magnitude. However, it is a common finding that decisions on green technology investments are much more sensitive to the former than the latter \citep{Jaffe:2005}. 
This is likely related to systematic deviations from EUT --- such as loss aversion, status quo bias and the salience effect \citep{Gillingham:2009}. In behavioural experiments, individuals were shown to weigh losses roughly twice as much as gains \citep{Greene:2011}.

In order to model the observed decision behaviour, we here adopt the positive framework of prospect theory. Specifically, we use a \emph{loss aversion value function} as proposed by \cite{Tversky:1992}, which conceptualises observed decision-making based on: (1) \emph{reference dependence}: gains and losses are defined relative to the status quo as a reference point, (2) \emph{loss aversion}: since losses weigh larger than gains, the function is steeper in the negative than in the positive domain --- captured by the parameter $\lambda$, (3) \emph{diminishing sensitivity}: gains and losses are subject to decreasing valuation, resulting in a s-shaped function --- expressed by the exponents $\alpha$ (for losses) and $\beta$ (for gains). Empirically, it is estimated that $\lambda=2.25$ and $\alpha = \beta = 0.88$ \citep{Tversky:1992}, which holds consistently in very different contexts \citep{Benartzi:1995}.

The decision criterion of level~3 is eq~\ref{eq:behavioural}, with $\sum_{t=0}^{b_{i}} E_{i}(\Delta B_{i, t}) =  \sum_{t=0}^{b_{i}}  \left[ E_{i}(p_{i, t}) * E_{i}(\Delta q_{i, t}) \right] $. 
\beq
NPB_{i} := - \lambda * (\Delta C_{i})^ {\beta} + \left[ \sum_{t=0}^{b_{i}} E_{i}(\Delta B_{i, t})\right] ^{\alpha}  = - 2.25 * (\Delta C_{i}) ^{0.88}  + \left[ \sum_{t=0}^{b_{i}} E_{i}(\Delta B_{i, t}) \right] ^{0.88}
\label{eq:behavioural}
\eeq

Note that eq~\ref{eq:behavioural} is based on the satisficing level, and therefore includes the explicit payback threshold $(b_{i})$. Given that individual decision-makers decide within the given organisational constraints of their firms, it is assumed that the long-term benefits beyond the cut-off point are not relevant for their individual decision utility.

\subsubsection*{(4) Ensemble of all decision types}

The different types of decision-making are not meant to be mutually exclusive. Rather, one type of firm might decide based on a decision rule similar to the NPV calculation in level 1, while others decide according to rules that are closer to the payback threshold of level 2. Finally, a remaining fraction of firms might be subject to the behavioural biases of individual decision-makers, as described best by level 3.

In reality, a mix of firms with different types of decision-making for the given investment is more likely than an exclusive prevalence of just one type of firm. To simulate different distributions of decision-making types among firms, we therefore construct an ensemble model that contains levels 1, 2 and 3. Each firm is randomly assigned to one level. The probability that an individual firm belongs to level 1, 2 or 3 is assumed to be exogenously given by $p(level=1)$, $p(level=2)$ and $p(level=3)$, which sum up to 1. 

Based on the results of \cite{Graham:2001} for decision-making of firms in the USA, we use the following default calibration:

\beq
 p(level=1) = 0.4;~p(level=2) = 0.3;~p(level=3) = 0.3
\label{eq:ensemble}
\eeq
 
In a Monte Carlo simulation, these probabilities will equal the fractions of firms assigned to the different levels. By means of numerical exploration, the probabilities can be adjusted to any value,  in order to analyse the potential effect of any population composition.

\section{Case study: electric motors in the USA \label{sect:CaseStudy}}

In this section, we parameterise the model using data for energy-efficient electric motors in the USA. We demonstrate what can be done with the model and how it can enhance the policy-maker's understanding of a green technology's rate of adoption. 
As an example, we simulate investment decisions on the different levels of decision-making and compare it to observed decisions after energy audits. In a situation which otherwise would have been given the un-instructive label of \emph{irrational behaviour}, the model allows for better understanding of the problem, and analysis of the likely effectiveness of hypothetical policies.

\subsection{Background information}

The International Energy Agency \citep{IEA:2011} estimates that electric motor driven systems (EMDS) account for 43--46\% of global electricity consumption, causing annual CO$_{2}$ emissions of roughly 6~Gt. At the same time, the IEA estimates that the energy efficiency of EMDS could be cost-effectively increased by 20--30\%. 

Throughout a potential lifetime of 20 years, electricity typically accounts for 90\% of a motor's life-cycle costs, compared to 1\% for the purchase price. As a result, even ``\emph{small gains in energy efficiency can be highly cost-effective}'' (p. 72) and should more than justify the initially higher investment costs (or even a premature replacement of existing motors). However, despite the apparent attractiveness of such green investments, the IEA considers their realisation as `\emph{difficult or impossible}' (p. 13) due to a variety of barriers --- for example higher initial costs $(\Delta C_{i})$ and short-term thinking of firms $(b_{i})$. Accordingly, investment decisions focus on low investment costs, and largely ignore the potential cost-savings. 

Within the model, this `\emph{efficiency gap}' corresponds to the engineering perspective of the benchmark level zero --- the technological level of decision-making. However, how are EMDS investments related to the other levels?

\subsection{Data and calibration}

The model is calibrated to data from the US Department of Energy's Industrial Assessment Centers (IAC) \citep{CAES:2015}, which provide energy audits for small and medium-sized manufacturing companies free of charge since 1976. Auditing teams perform in-depth assessments of factories. These consist of surveys, engineering measurements and a two-day site visit. The auditors then provide the respective company with a detailed investment analysis of projects that were identified as cost-effective. After six months, auditors perform follow-up interviews on which projects are implemented (ibid.). The key data of all project recommendations is consistently reported in a public database, including: estimated investment costs $(\Delta C_{i})$, estimated annual energy savings $(\Delta q_{i})$, electricity prices $(p_{i})$ and the final implementation status. 

We focus on audits between 2008 and 2013, assuming that the motor technology throughout this period is roughly comparable. After removing outliers (with reported investment costs being either zero or unrealistically high), this provides information on 275 recommendations for the `\emph{use (of the) most efficient type of electric motors}'. Note that this only refers to a motor's complete replacement (reconfigurations are listed separately), allowing a treatment as a roughly homogenous investment. 

Descriptive statistics for all variables are summarised in table~\ref{tab:data}. The data on investment costs and quantities is transformed into relative values (\$ per kWh saved and kWh saved per \$). For illustrative purposes, this data is then scaled to the median project size (as measured by kWh saved). All stated costs and prices are expressed in 2013-US-Dollars.

\begin{table}\footnotesize
\begin{center}
		\begin{tabular*}{1\columnwidth}{@{\extracolsep{\fill}} l |  l | l | l | l | l | l | l }
			
Variable & Unit	& Mean	& Median	& Min.	& 95\% perc. &	Max.	& Std. dev.\\ 
\hline
\hline
\emph{Absolute:} & & & & & & & \\
\hline
$\Delta C_{i}$&	\$	&32 258	&8 224	&264 &	115 640	&993 790	&101 190 \\
$\Delta q_{i}$ & $\Delta kWh/y$ &	178 250	&46 050	&999	&708 940	&5 645 400	&496 380 \\
$p_{i}$ & \$/kWh&	0.073	&0.070&	0.020&	0.130&	0.220&	0.031\\
$\Delta B_{i}$ & \$/y	& 10 293 &	3 319 &	82	&41 693	&233 950	&23 329 \\
payback time	& years&	3.34&	2.65&	0.15&	8.83&	16.41&	2.57\\
\hline
\emph{Relative:} & & & & & & & \\
\hline
$C_{i}/\Delta q_{i}$ & $\$/\Delta kWh$ &	0.23&	0.19&	0.03&	0.54&	0.79&	0.16 \\
$\Delta q_{i}/C_{i}$ & $\Delta kWh/\$$ & 7.56	&65.26&	1.26&	22.44&	34.08&	6.37\\
\hline
\emph{Median motor:} & & & & & & & \\
\hline
$C_{i}$ & \$	& 10 311 &	8 685&	1 341&	24 527&	36 256&	7 270\\
$\Delta q_{i}$ & $\Delta kWh/y$ & 62 205&	43 280&	10 368&	184 550&	280 290&	52 361 \\
\hline		
		\end{tabular*}
	\caption{Summary statistics of motor recommendations (absolute, relative, median).}
	\label{tab:data}
\end{center}
\end{table}

\begin{figure}
	\begin{minipage}[t]{.5\columnwidth}
		\begin{center}
			\includegraphics[width=1\columnwidth]{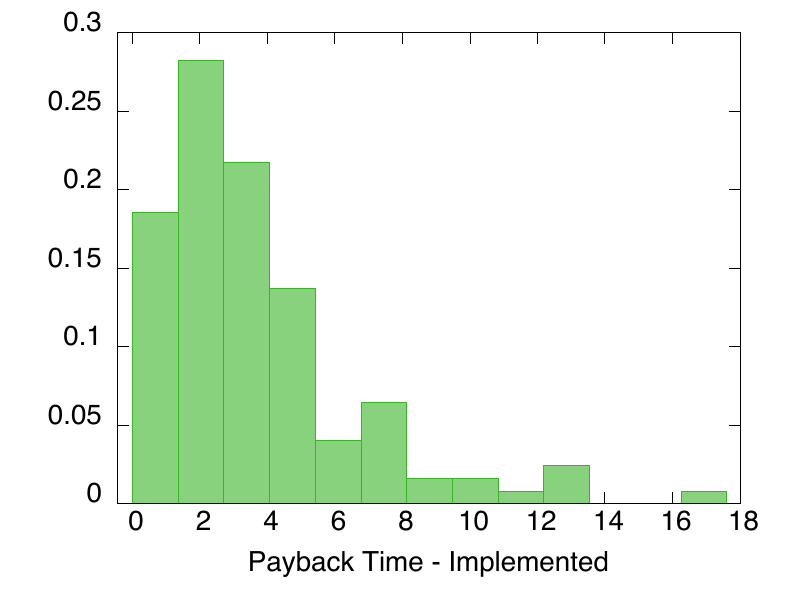}
		\end{center}
	\end{minipage}
	\begin{minipage}[t]{.5\columnwidth}
		\begin{center}
		        \includegraphics[width=1\columnwidth]{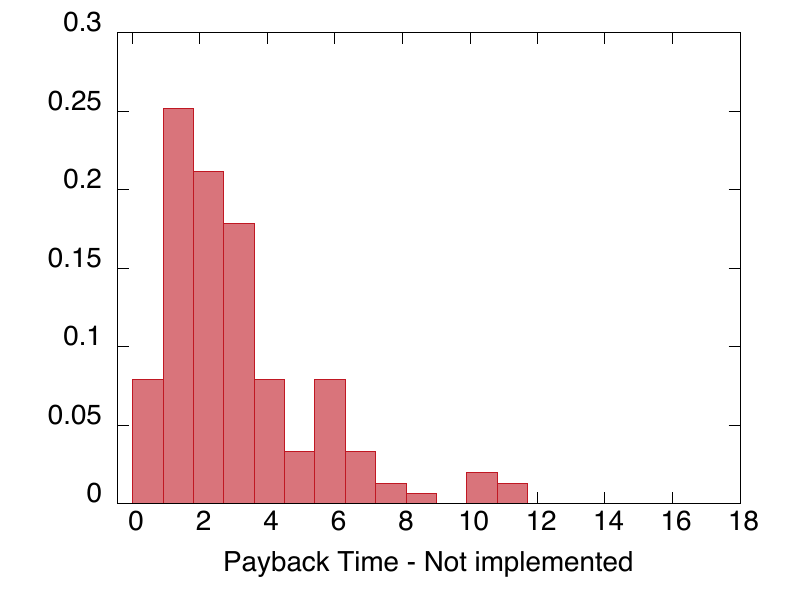}
		\end{center}
	\end{minipage}
	\caption{Relative frequencies of project payback times, grouped by implementation status (green: implemented, red: not implemented).}
	\label{fig:Figure2}
\end{figure}

First, the data reveal that there is considerable heterogeneity with respect to estimated investment costs, annual energy savings and electricity prices (figure~\ref{fig:Figure3}~a.). Second, calculated payback times are always shorter than an electric motor's expected lifetime, and shorter than 8.83 years for 95\% of all projects. Therefore, the investment seems indeed to be profitable for all firms. Despite these estimations, the observed implementation rate is as low as 45\%. This seems to be unrelated to the considered economic fundamentals: by graphical inspection, project payback times do not differ by implementation status (see figure~\ref{fig:Figure2}). This indicates the relevance of other levels of decision-making.

The model is calibrated to the IAC data for $\Delta C_{i}$, $\Delta q_{i}$ and $p_{i}$ based on the estimations of median-scaled motor projects (see table~\ref{tab:calibration}). To avoid having to specify a particular price path, $p_{i}$ is assumed to be constant throughout an investment's lifetime (so is $\Delta q_{i}$). Calibrations of discount rates ($r_{i}$), lifetimes ($n_{i}$), payback thresholds ($b_{i}$) and the prospect theory loss aversion function ($\lambda$, $\alpha$, $\beta$) are based on literature estimates. 

To allow for Monte Carlo simulations, the empirical data ($\Delta C, \Delta q, p$) is approximated by theoretical distributions. Graphical comparisons with different distribution types as well as bootstrapped Cullen and Frey graphs (plotting the kurtosis against the squared skewness) indicate that the best fit is given by Weibull distributions. Parameter fitting was done by maximum likelihood estimation. 

The investment's expected lifetime ($E_{i}(n_{i})$), cost of capital ($r_{i}$) and payback thresholds ($b_{i}$) are approximated by Normal distributions. 
$b_{i}$ is truncated between a minimum of 1 year and a maximum of 5 years \citep[based on][]{Anderson:2004}.

\begin{table}\footnotesize
\begin{center}
		\begin{tabular*}{1\columnwidth}{@{\extracolsep{\fill}} l |  l | l | l | l}
			
Level	&Variable	&Distribution&	Parameter values	&Source\\ 
\hline
\hline
1	& $\Delta C$ &	Weibull& 	shape=1.51; scale=11~493.28 & fitted to data from the\\
\cline{2-4}
	& $p$ &	Weibull& 	shape=2.46; scale=0.08 & \cite{CAES:2015}\\
    \cline{2-4}
    	& $\Delta q$ &	Weibull& 	shape=1.34; scale=68~426.27 & \\
\cline{2-5}	
 & $r$& Normal	& $\mu$= 8\%; $\sigma$ = 3\%; min=0\% & \cite{Gilchrist:2013}\\
 \cline{2-5}	
 & $n$ &	Normal	& $\mu$=15; $\sigma$ =3; min=0	& \cite{IEA:2011}\\
  \cline{2-5}
 & $\gamma_{i}$ &	--- &	1 & \emph{benchmark}\\
 \hline
2 & $b$ & Normal & $\mu$=2; $\sigma$ =1; min=1; max=5 & \cite{Anderson:2004}\\
\hline
3 & $\lambda$ & --- & 2.25 &  \\
\cline{2-4}
   & $\alpha$ & ---& 0.88 & \cite{Tversky:1992}\\
\cline{2-4}
   & $\beta$ & --- & 0.88 &  \\
   \hline
   Ensemble & $p(level)$ & Discrete empirical & p(level=1)=0.4; p(level=2)=0.3; p(level=3)=0.3 &  \cite{Graham:2001}  \\
   \hline
		\end{tabular*}
	\caption{Calibration of model parameters to electric motor project recommendations.}
	\label{tab:calibration}
\end{center}
\end{table}

\subsection{Simulation results}

\begin{figure}
	\begin{minipage}[t]{.5\columnwidth}
		\begin{center}
			\includegraphics[width=1\columnwidth]{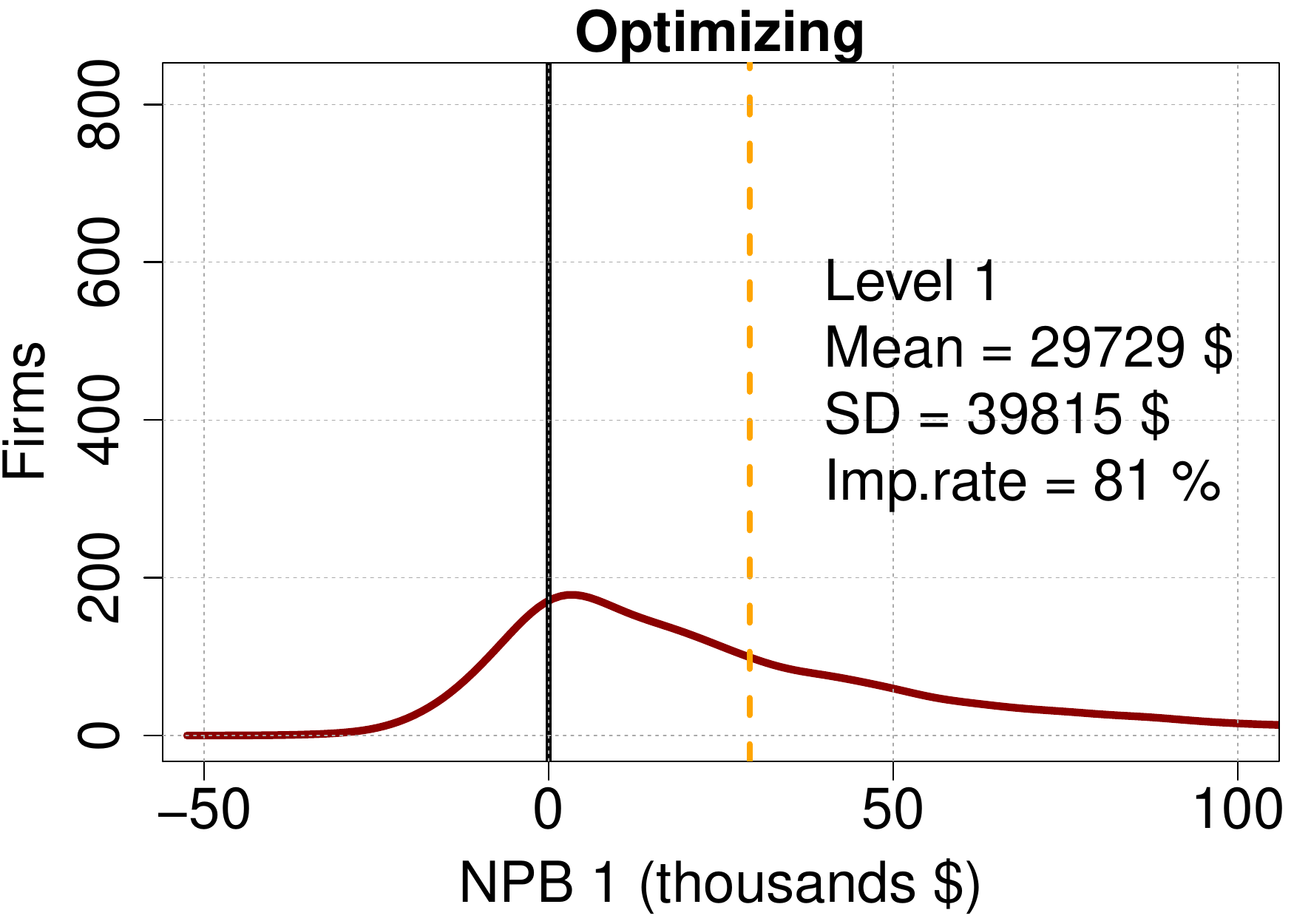}
		\end{center}
	\end{minipage}
	\begin{minipage}[t]{.5\columnwidth}
		\begin{center}
		        \includegraphics[width=1\columnwidth]{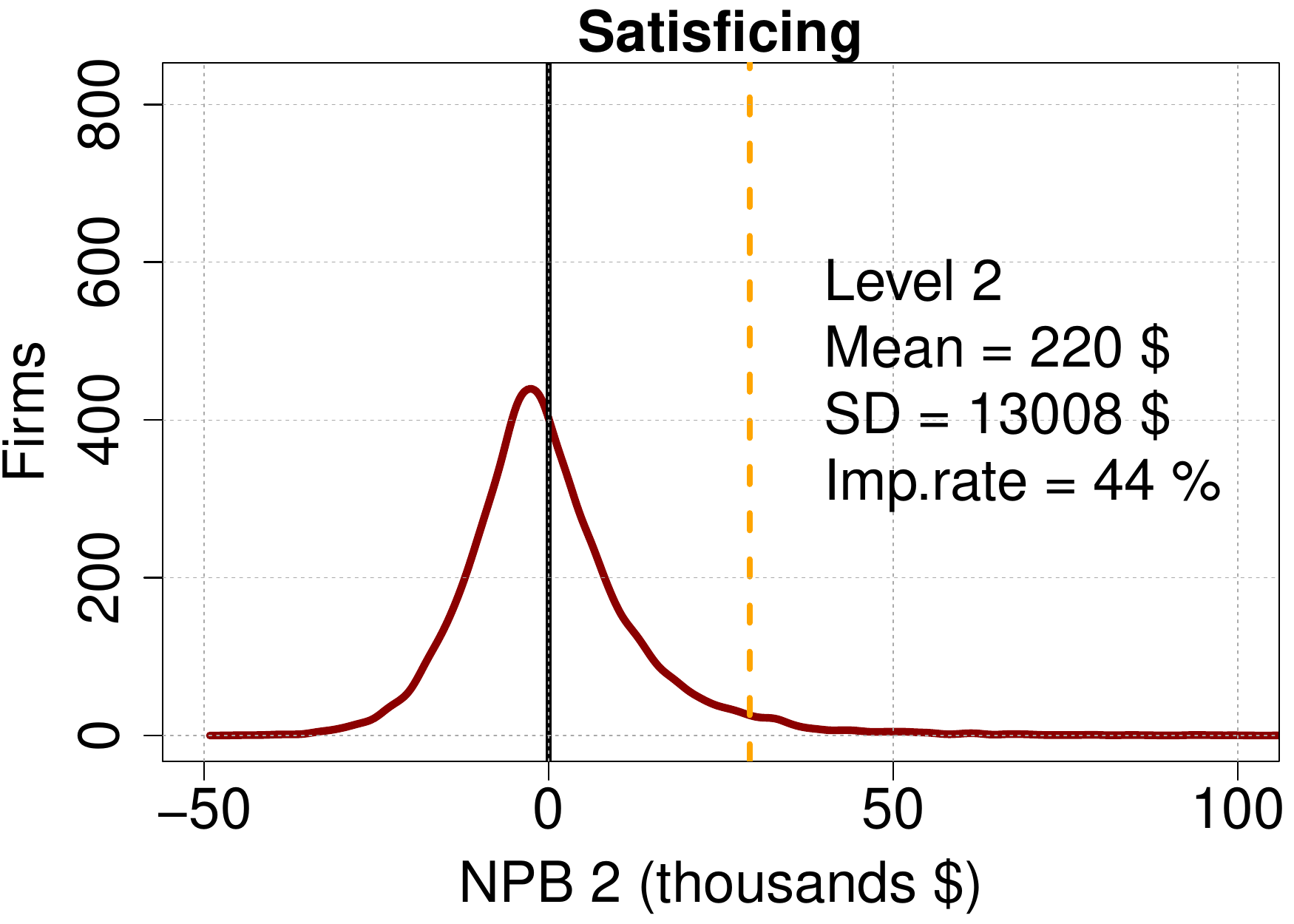}
		\end{center}
	\end{minipage}
		\begin{minipage}[t]{.5\columnwidth}
		\begin{center}
			\includegraphics[width=1\columnwidth]{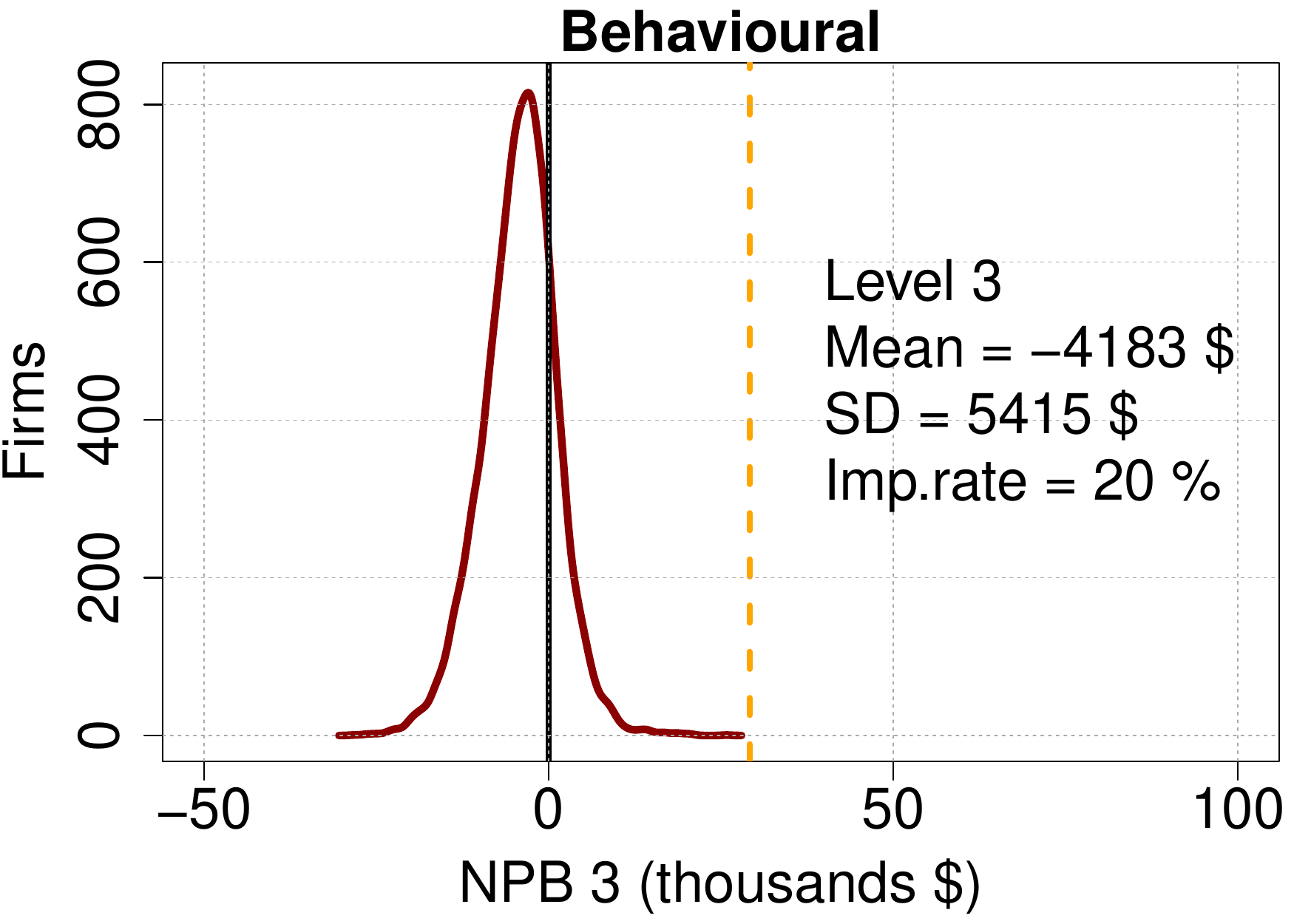}
		\end{center}
	\end{minipage}
	\begin{minipage}[t]{.5\columnwidth}
		\begin{center}
		        \includegraphics[width=1\columnwidth]{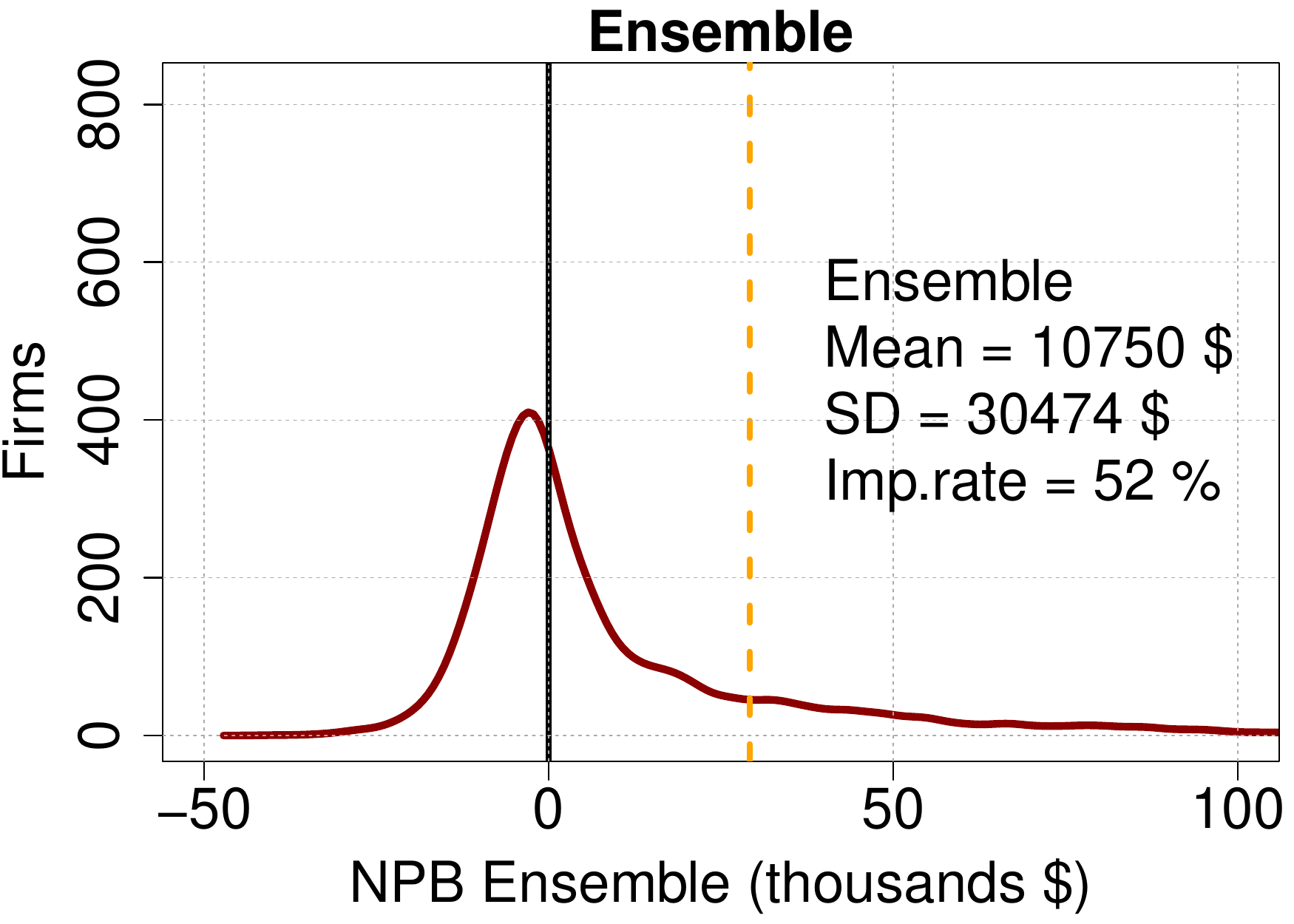}
		\end{center}
	\end{minipage}
	\begin{minipage}[t]{1\columnwidth}
		\begin{center}
		        \includegraphics[width=0.45\columnwidth]{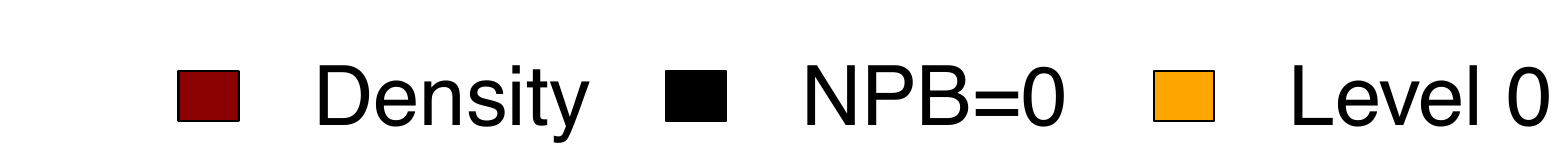}
		\end{center}
	\end{minipage}
	\caption{Distributions of simulated NPV/NPB for the modelled levels of decision-making (representations of evaluation methods) 1 to 3, respectively "Optimizing", "Satisficing" and "Behavioural" (see text), as well as the model ensemble (mixture of levels). The density estimations are shown with the break even line at zero, as well as the point estimate of level 0 as a dashed line.}
	\label{fig:Figure3}
\end{figure}

To estimate the share of firms that would invest in a more efficient motor according to the different levels of decision-making (referred to as the implementation rate), a Monte Carlo simulation of the calibrated model was run with 10~000 trials. This enables to combine (or convolve) chains of probability distributions across levels (equivalent to formal mathematical convolutions). Resulting implementation rates and summary statistics of the NPV/NPB distributions are summarised in table~\ref{tab:simulationresults}. Density estimations of the distributions at different levels are depicted in figure~\ref{fig:Figure3}.

\begin{table}[t]\footnotesize
\begin{center}
		\begin{tabular*}{1\columnwidth}{@{\extracolsep{\fill}} l |  c | c | c | c |  c | c}
			
Level	& $NPV>0$	&Mean	&Median	&Minimum	&Maximum	&Std. dev.\\ 
\hline
\hline
 0 --- Technological	& 100.0\% & 30~144 &--- &--- & --- & 0 \\
 1 --- Optimizing &	81\%	&29~729	&18~711	&$-$39~068&	476~723	&39~815 \\
 2 --- Satisficing	& 44\%	&220	& $-$1~391	&$-$44~887&	124~314&	13~008\\
 3 --- Behavioural &	20\%	&$-$4~183	&$-$3~907	&$-$28~290	&25~848	&5~415 \\
 E --- Ensemble &	52\%	&10~750	&499	&$-$40~480	&335~300	& 30~474 \\
 \hline
		\end{tabular*}
	\caption{Model simulations for levels 0 to 3 and the model ensemble --- Implementation rates (in \%) and summary statistics of the simulated NPV/NPB (in US\$).}
	\label{tab:simulationresults}
\end{center}
\end{table}

First, the model simulation for the optimizing level of decision-making is consistent with the IEA analysis: the investment in a more energy-efficient motor appears to be profitable for a large majority of 81\% of firms. However, it proves to be unprofitable for the remaining 19\%. Although the investment is highly profitable on average (with a mean NPV of 29~729\$), the considered characteristics among firms are highly heterogeneous. This implies that a fraction of them is characterised by such combinations of high costs and low benefits that make it perfectly reasonable to dismiss the motor investment.

Second, the simulated implementation rate decreases for the two other levels --- from 81\% on the optimizing level to 44\% on the satisficing level and 20\% on the behavioural level. In other words, the distributions of simulated NPBs shift to the left. At level~3, only one in five firms is predicted to invest, although it is seen as cost-effective for four out of five firms from an optimizing perspective. This has to be contemplated in the context of a relatively well-proven and risk-free green investment, proposed to firms in the relatively credible form of government-sponsored energy audits. 

Third, the input data's combined heterogeneity translates into even more heterogeneous investment perceptions by firms, when evaluated in form of NPVs and NPBs. From the optimizing perspective of level 1, simulated NPVs range from $-$39~068\$ to 476~723\$, with a standard deviation of 39~815\$ --- due to the investment's heterogeneous economic fundamentals. The variance decreases along with the levels, which can be explained by two effects. From level 1 to 2, it decreases along with the considered time frame. This is because a large part of the variance can be attributed to heterogeneous annual payoffs, and their relevant magnitude is much smaller when assuming a payback criterion (as on levels 2 and 3) instead of a full lifetime assessment (as on level 1). On level 3, the prospect theory loss aversion function has two implications. The overweighting ($\lambda$) of losses associated with the upfront costs decreases perceived NPBs. At the same time,  the decreasing marginal utility of both losses and gains ($\alpha$ and $\beta$) imply a less heterogeneous perception of costs and benefits, further decreasing the variance.

Finally, the IAC data allows a comparison with the observed implementation rate. 45\% of firms report that they have implemented the recommended efficiency improvement, matching our result for the satisficing level~2 (44\%). The large gap between this rate and the 81\% predicted by level~1 (the actual \emph{energy efficiency gap}) can be interpreted as evidence for organisational and behavioural barriers, e.g. in the form of short-term investment planning and loss aversion. However, from a policy perspective, this divergence could as well be interpreted as evidence for the audits' effectiveness: if we are to believe our model, without it, the implementation rate could have been as low as 20\%, the model's estimation for the behavioural level~3. By offering technological assistance and pointing out profitable projects, the audits are very likely to have \emph{shifted} the decision from the behavioural to the satisficing level.\footnote{However this would remain to be proven, which cannot be done with the current data, as it would require similar data in a counterfactual situation without energy audits, but the audits are the actual source of the data.} 
As a consequence, the investment decision might be less prone to behavioural factors than it would have been without the audit. 

However, since the input data for many parameters was estimated and randomly assigned to hypothetical firms, the model output can only be interpreted in aggregate. We don't know if firms really decide based on a decision-making procedure similar to level 2, of which the prediction roughly matches the observed outcome. From a macro-perspective, it might as well be that around 40\% of firms are of the optimizing decision type of level 1 (as found by \cite{Graham:2001}), while the remaining ones are closer to the satisficing or behavioural levels of decision-making. This would result in an implementation rate of 52\%, as demonstrated by the ensemble model. Neither do we know which share of firms does actually decide based on level 3 (or might have decided in the absence of any audit), and if the audit indeed shifted the level of decision-making for some firms. The fundamental problem here is similar to `\emph{Schr\"odinger's cat}': we can't measure firms' behaviour in the absence of an audit, while the audit itself most likely influences what we want to measure.

\subsection{Sensitivity analysis\label{sect:Sens}}

\begin{figure}
	\begin{minipage}[t]{.5\columnwidth}
		\begin{center}
			\includegraphics[width=1\columnwidth]{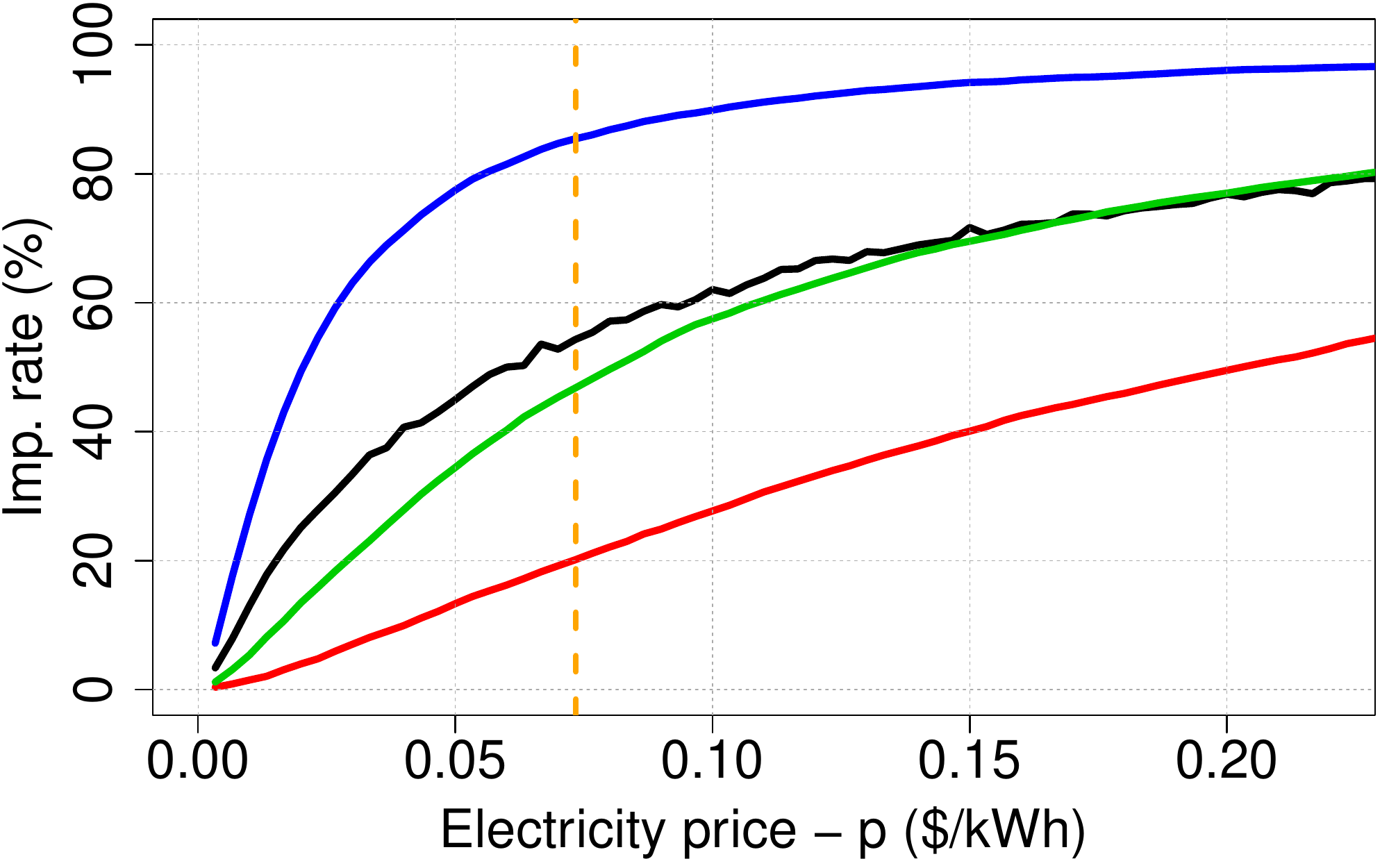}
		\end{center}
	\end{minipage}
	\begin{minipage}[t]{.5\columnwidth}
		\begin{center}
		        \includegraphics[width=1\columnwidth]{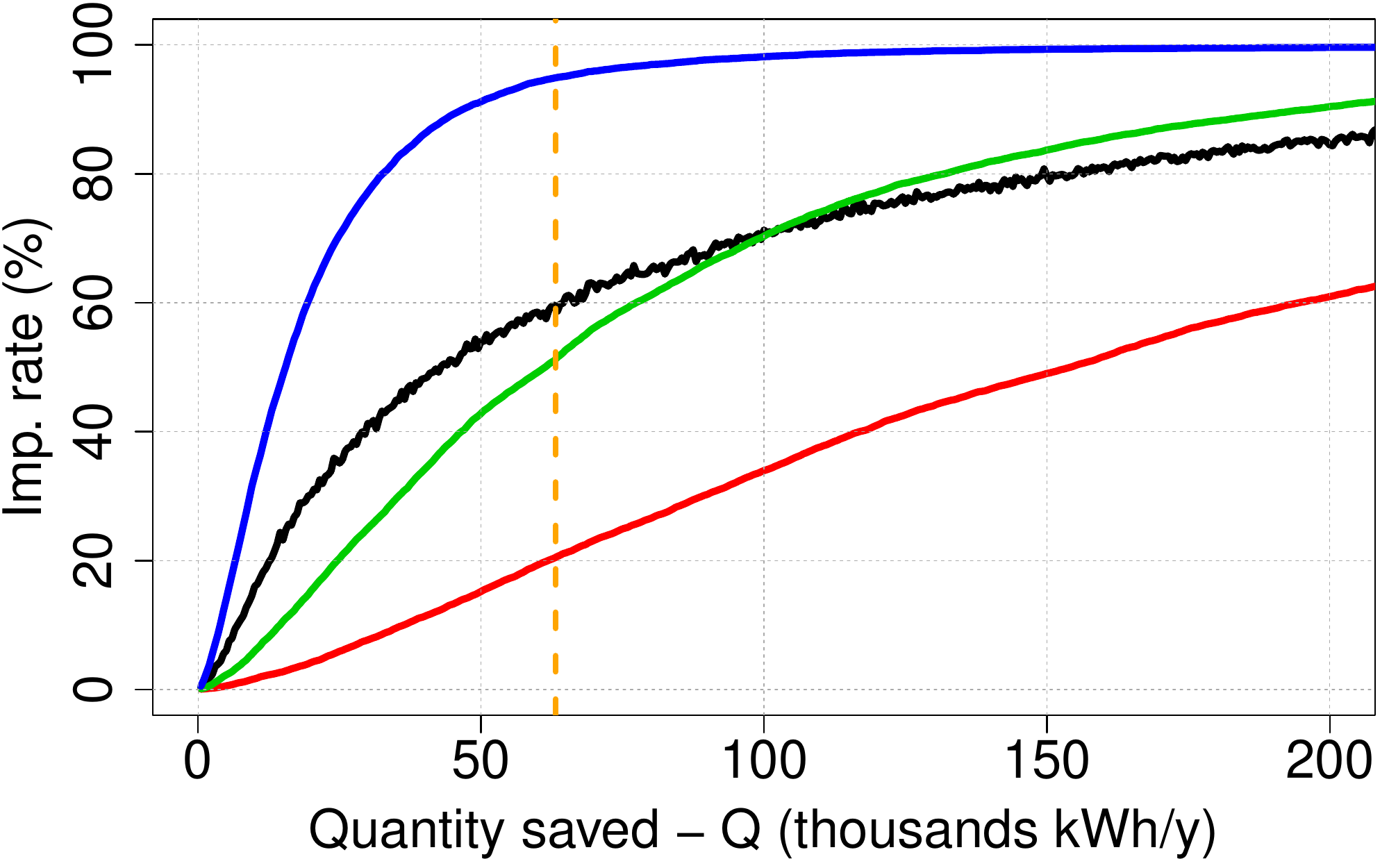}
		\end{center}
	\end{minipage}
		\begin{minipage}[t]{.5\columnwidth}
		\begin{center}
			\includegraphics[width=1\columnwidth]{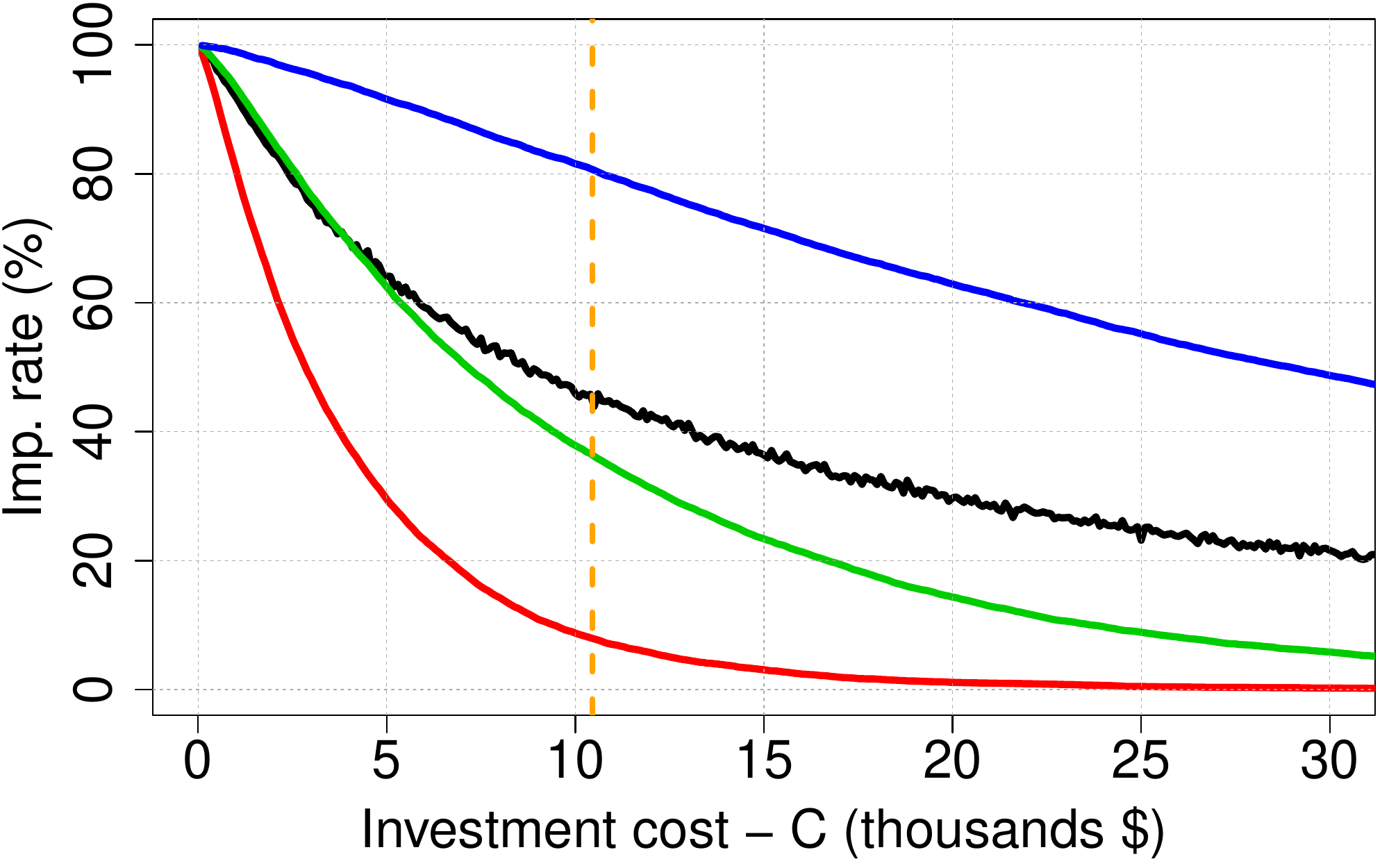}
		\end{center}
	\end{minipage}
	\begin{minipage}[t]{.5\columnwidth}
		\begin{center}
		        \includegraphics[width=1\columnwidth]{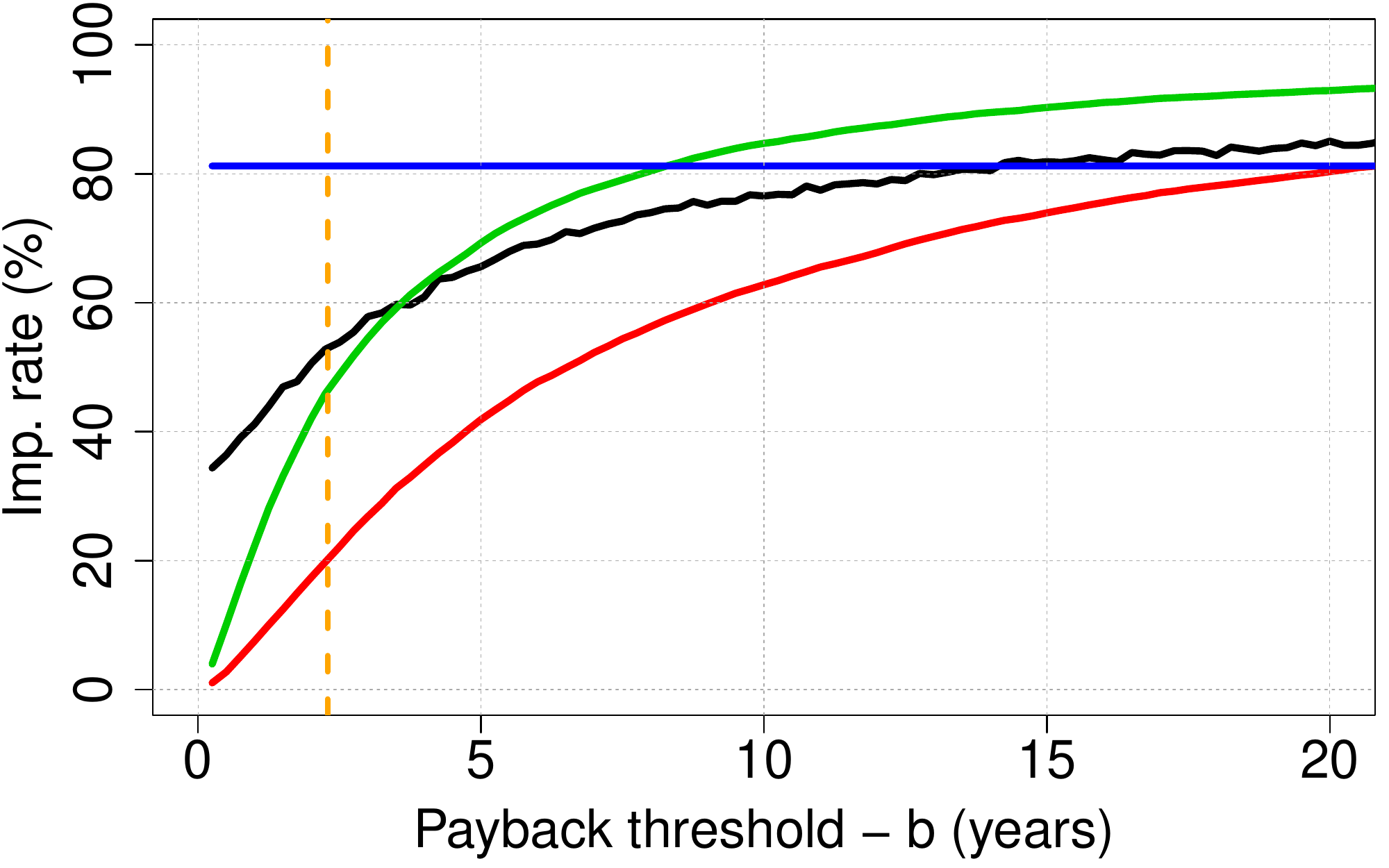}
		\end{center}
	\end{minipage}
			\begin{minipage}[t]{.5\columnwidth}
		\begin{center}
			\includegraphics[width=1\columnwidth]{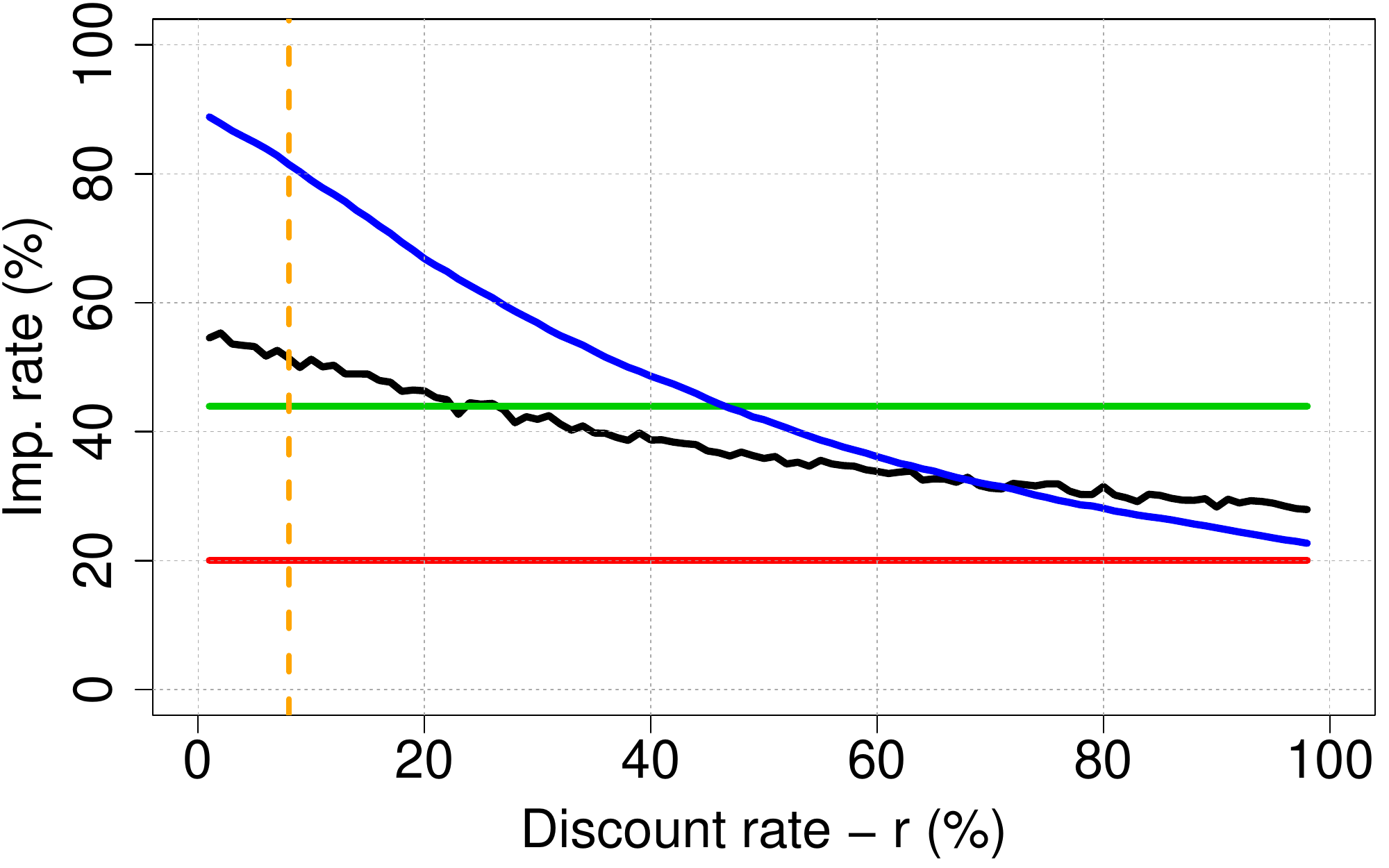}
		\end{center}
	\end{minipage}
	\begin{minipage}[t]{.5\columnwidth}
		\begin{center}
		        \includegraphics[width=1\columnwidth]{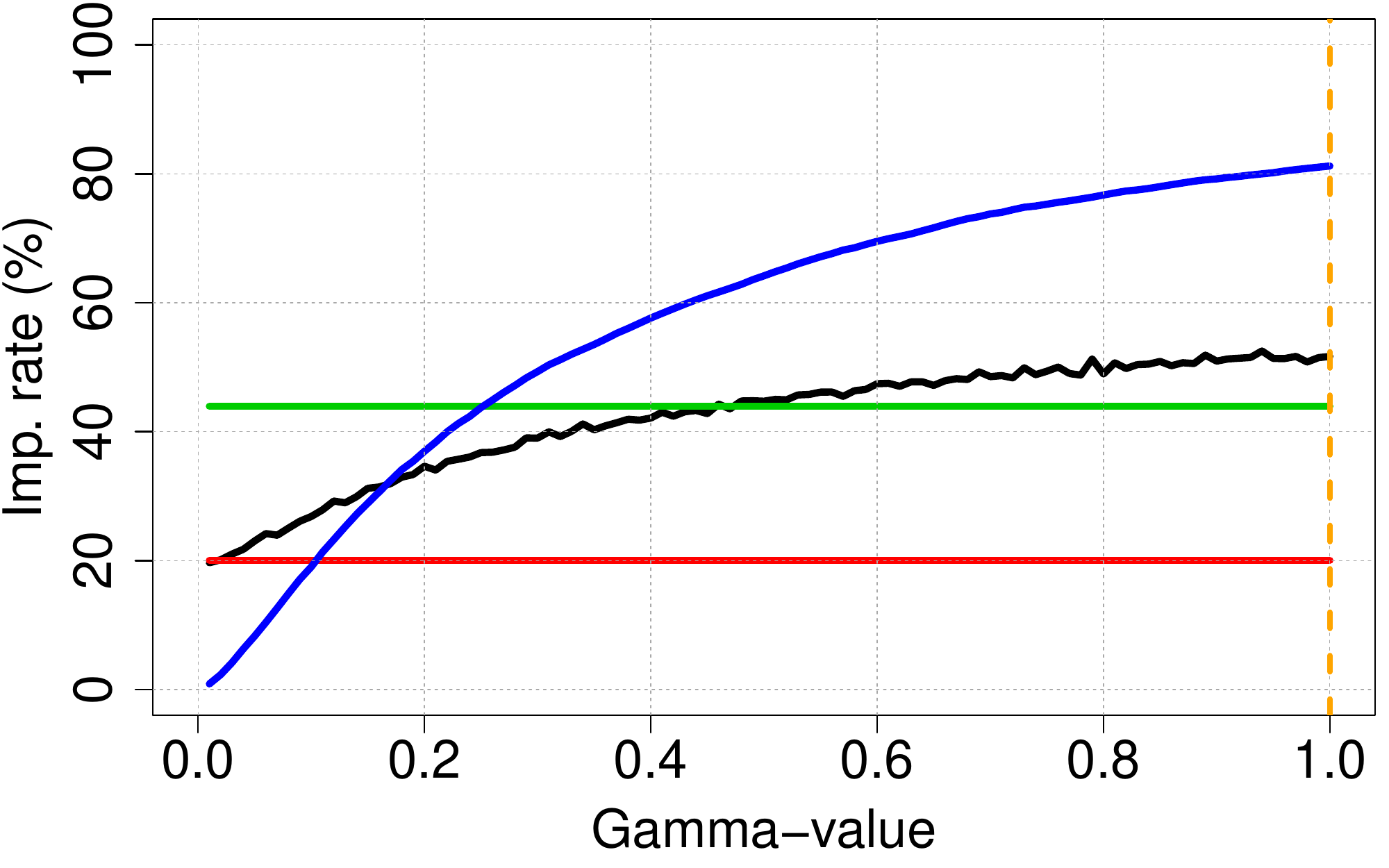}
		\end{center}
	\end{minipage}
	\begin{minipage}[t]{1\columnwidth}
		\begin{center}
		        \includegraphics[width=1\columnwidth]{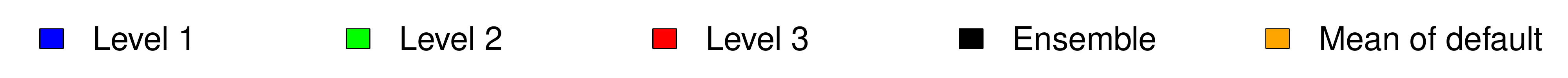}
		\end{center}
	\end{minipage}
	\caption{Sensitivity of the implementation rate (on different levels) against the main model parameters: electricity price (p), quantity saved (Q), investment cost (C), payback threshold (b), discount rate (r), gamma value. Colours represent the different levels of decision-making (see text).}
	\label{fig:Figure4}
\end{figure}

To investigate how individual model parameter affect outcomes, one parameter each is replaced with a range of point estimates. Figure~\ref{fig:Figure4} shows the simulated impacts on implementation rates.

The positive impact of the quantity saved ($\Delta q$) and the electricity price ($p$) is largest on level 1, on which the decision criterion accounts for benefits throughout the entire lifetime. It is lower on levels 2 and 3, on which benefits are underweighted relative to costs. For the same reason, the negative impact of higher investment costs ($C$) is largest on level 3, and lowest on level 1. 

The sensitivity towards the discount rate ($r$) and the parameter $\gamma$ allow an implicit estimation of the effect that imperfect information and risk aversion need to have on firms' valuation of benefits for an explanation based on pure optimizing: which parameter values are consistent with the observed implementation rate of 45\%, assuming that all firms were deciding based on level 1? The implicit discount rate is as high as 45\%, while the implicit level of $\gamma$ is as low as 0.26 --- meaning that firms would value potential cost savings at only 26\% of their value as promised by energy auditors. However, across all project categories within the IAC database, only 1.8\% of firms state that `\emph{risk of problem with product/equipment}' was a reason for non-adoption \citep{Anderson:2004}. An explanation purely based on high risk-perceptions and imperfect information therefore seems unlikely, and would at the very least be insufficient.

All parameters show a decreasing marginal impact --- which is relevant for the design of policies. While the simulated changes in average NPBs are always linear, they translate into a decreasing impact on firms' investment decisions --- consistent with the diffusion pattern that results from the distribution of heterogeneous adopters (see figure~\ref{fig:Figure1}). The reason is as follows: for more extreme parameter values, the linear increase (or decrease) in the mean NPB takes place along the thinner ends of NPB distributions, so that it encourages (or discourages) less and less firms to invest. Therefore, the higher the initial implementation rate, the lower the sensitivity towards $p$, and the higher the sensitivity towards $C$.

\subsection{Evaluating the effectiveness of hypothetical policies on implementation rates\label{sect:Scen}}

Here, policies are introduced in the model, aiming at increased implementation rates. From the perspective of welfare economics, there are at least two rationales for doing so: uninternalised externalities of electricity production and investment inefficiencies in the form of barriers to green technology investments \citep{Allcott:2012}.

\subsubsection*{Uninternalized externalities: electricity tax}

To account for negative externalities (e.g. CO$_{2}$ emissions), a Pigouvian tax on energy could be introduced, aiming at lower consumption. For manufacturing firms, this is equivalent to increasing electricity prices by $\Delta p_{i}$, thereby increasing the investment's annual payoff ($E_{i}(\Delta B_{i, t})$). From the optimizing perspective of level~1, a tax should induce fully rational firms to substitute away from electricity, and increase investments in green technologies.\newline 

{\bf Model input:} An electricity tax ($tax:=+\Delta p_{i}$) is simulated as a relative increase in a firm's individual electricity price per kWh ($p_{i}$) by a tax rate $t$, so that the electricity price after tax equals $(1+t)*p_{i}$. 

\subsubsection*{Investment inefficiencies: subsidies}

When firms' decisions focus on upfront costs relative to future payoffs, this constitutes an investment inefficiency. Reduced investment costs then promise a more effective leverage on decisions, and capital subsidies $(-\Delta C_{i})$ could be socially beneficial \citep{Allcott:2012}. \newline 

{\bf Model input:}  A subsidy scheme ($subsidy=:-\Delta C_{i}$) is simulated as a relative reduction in firms' upfront investment costs ($C_{i}$) by a subsidy rate $s$, so that the investment cost after the subsidy equal $(1-s)*C_{i}$.  

\subsubsection*{The impact of financial incentives}

\begin{figure}
	\begin{minipage}[t]{.5\columnwidth}
		\begin{center}
			\includegraphics[width=1\columnwidth]{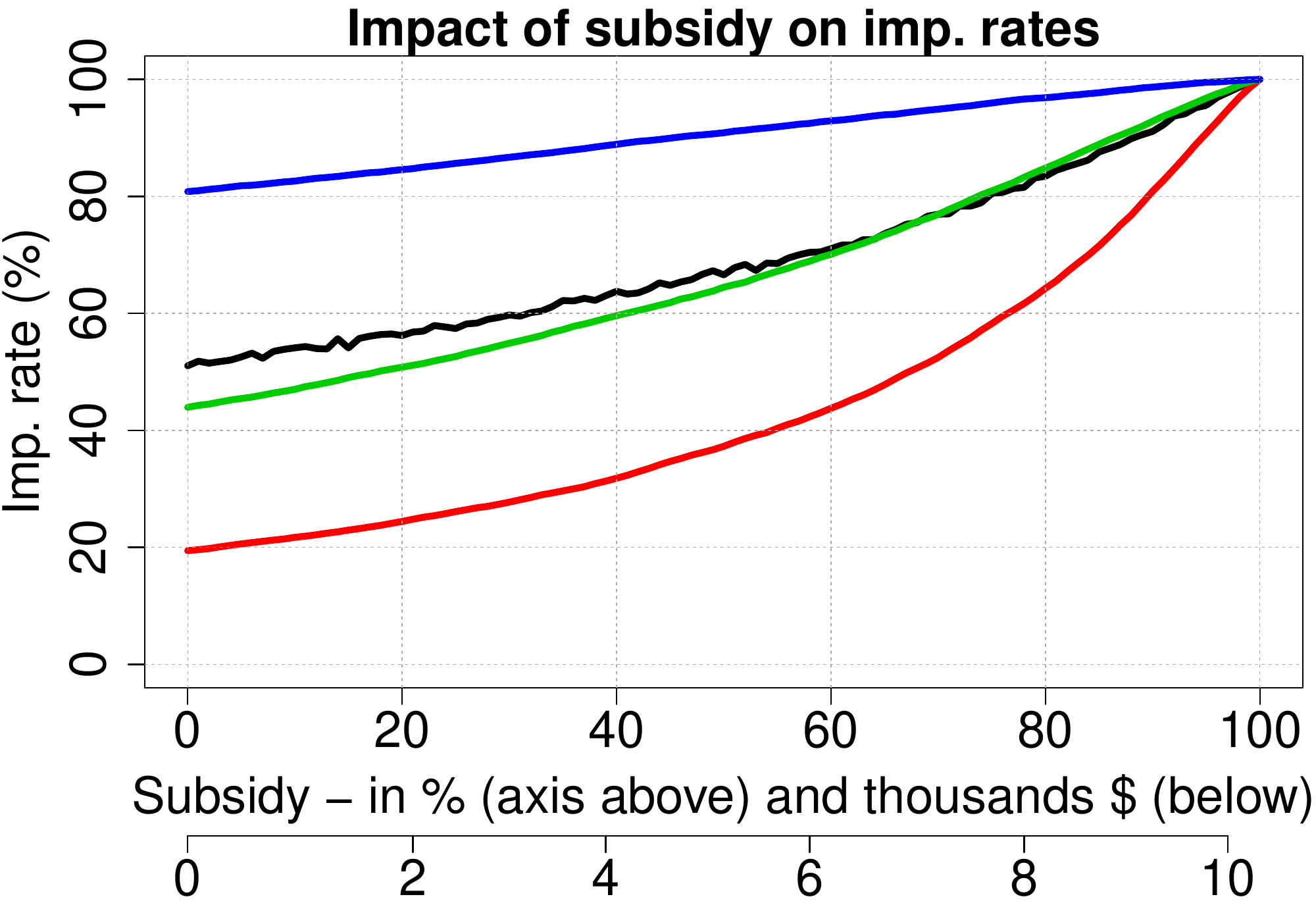}
		\end{center}
	\end{minipage}
	\begin{minipage}[t]{.5\columnwidth}
		\begin{center}
		        \includegraphics[width=1\columnwidth]{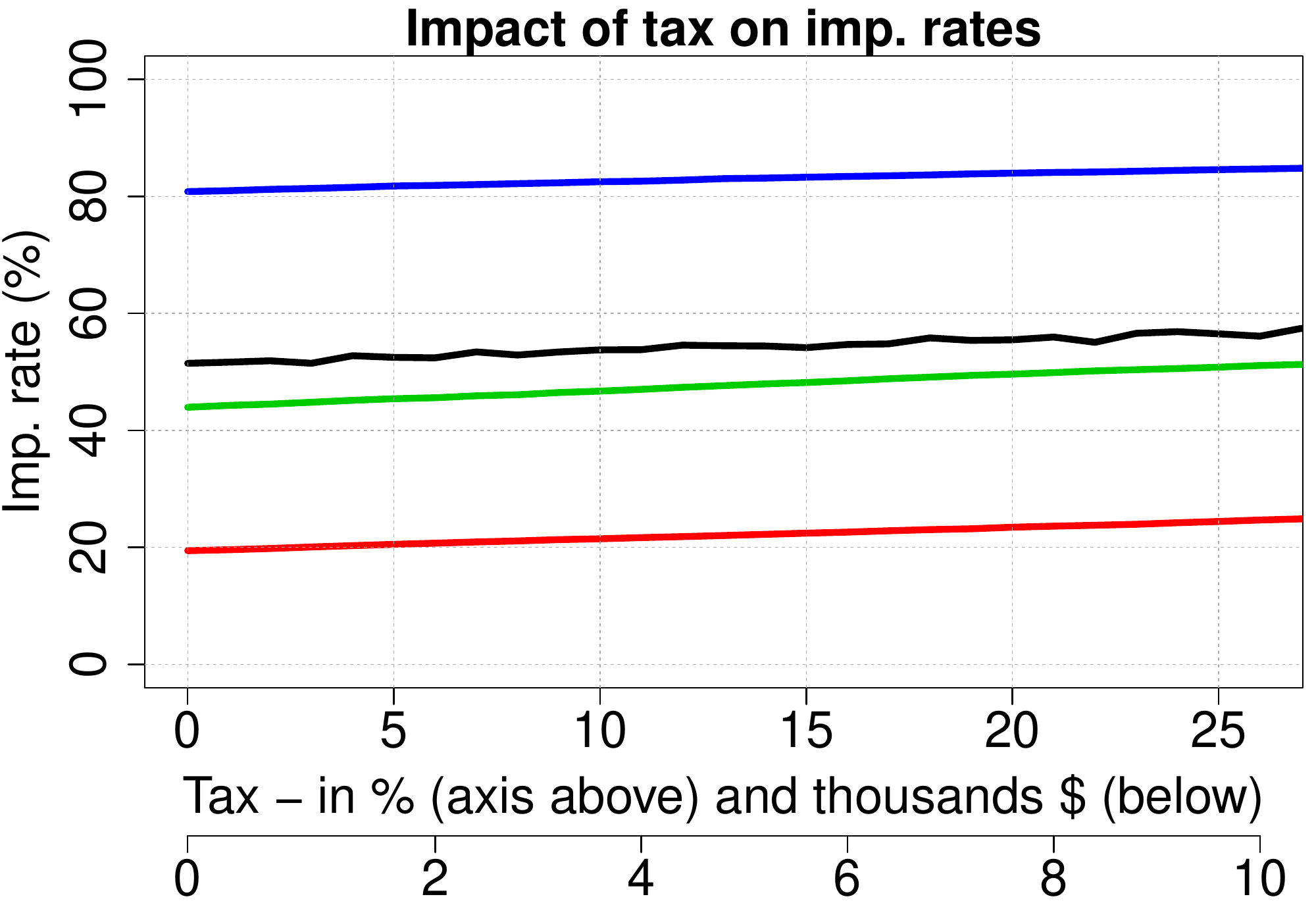}
		\end{center}
	\end{minipage}
		\begin{minipage}[t]{.5\columnwidth}
		\begin{center}
			\includegraphics[width=1\columnwidth]{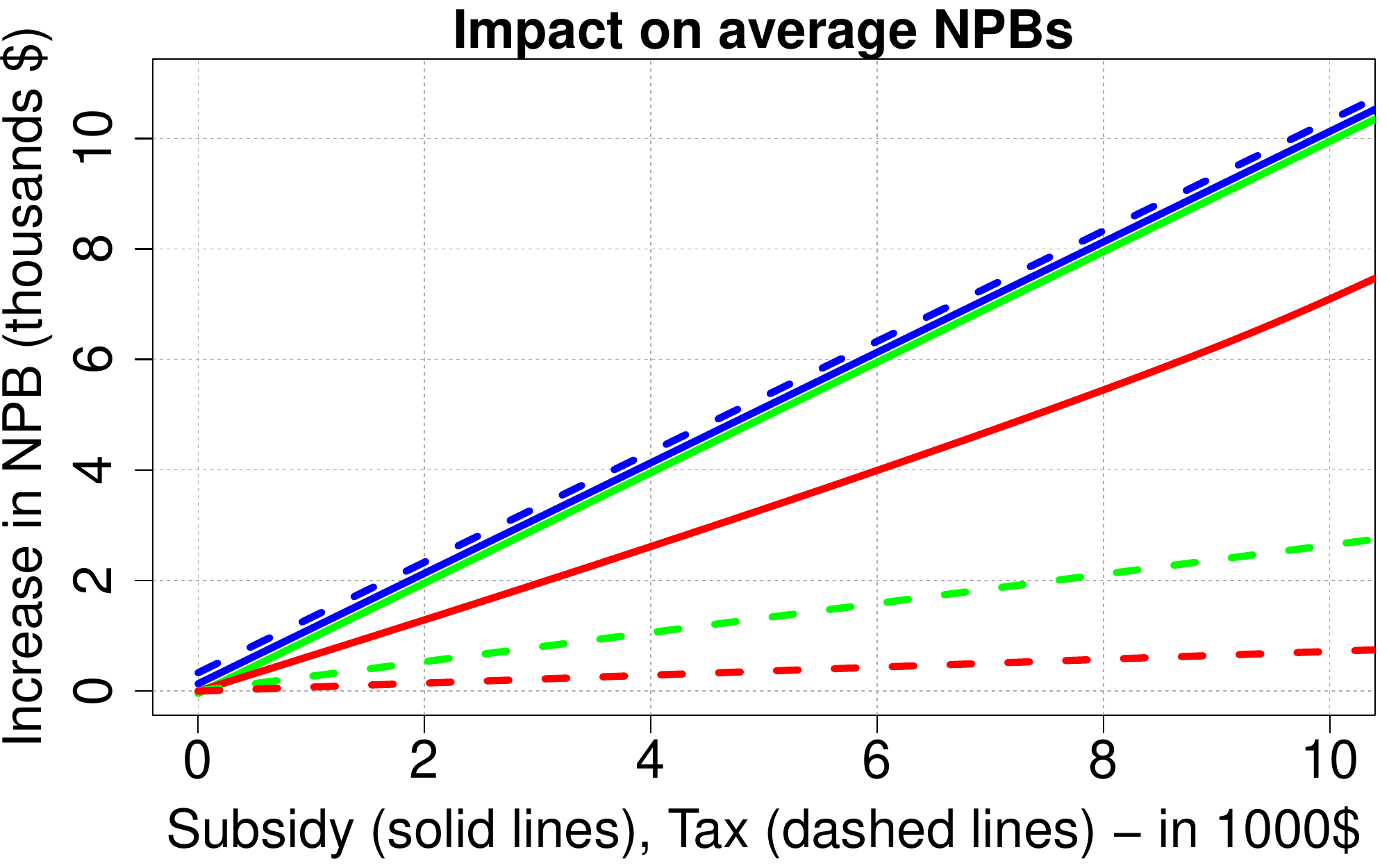}
		\end{center}
	\end{minipage}
	\begin{minipage}[t]{.5\columnwidth}
		\begin{center}
		        \includegraphics[width=1\columnwidth]{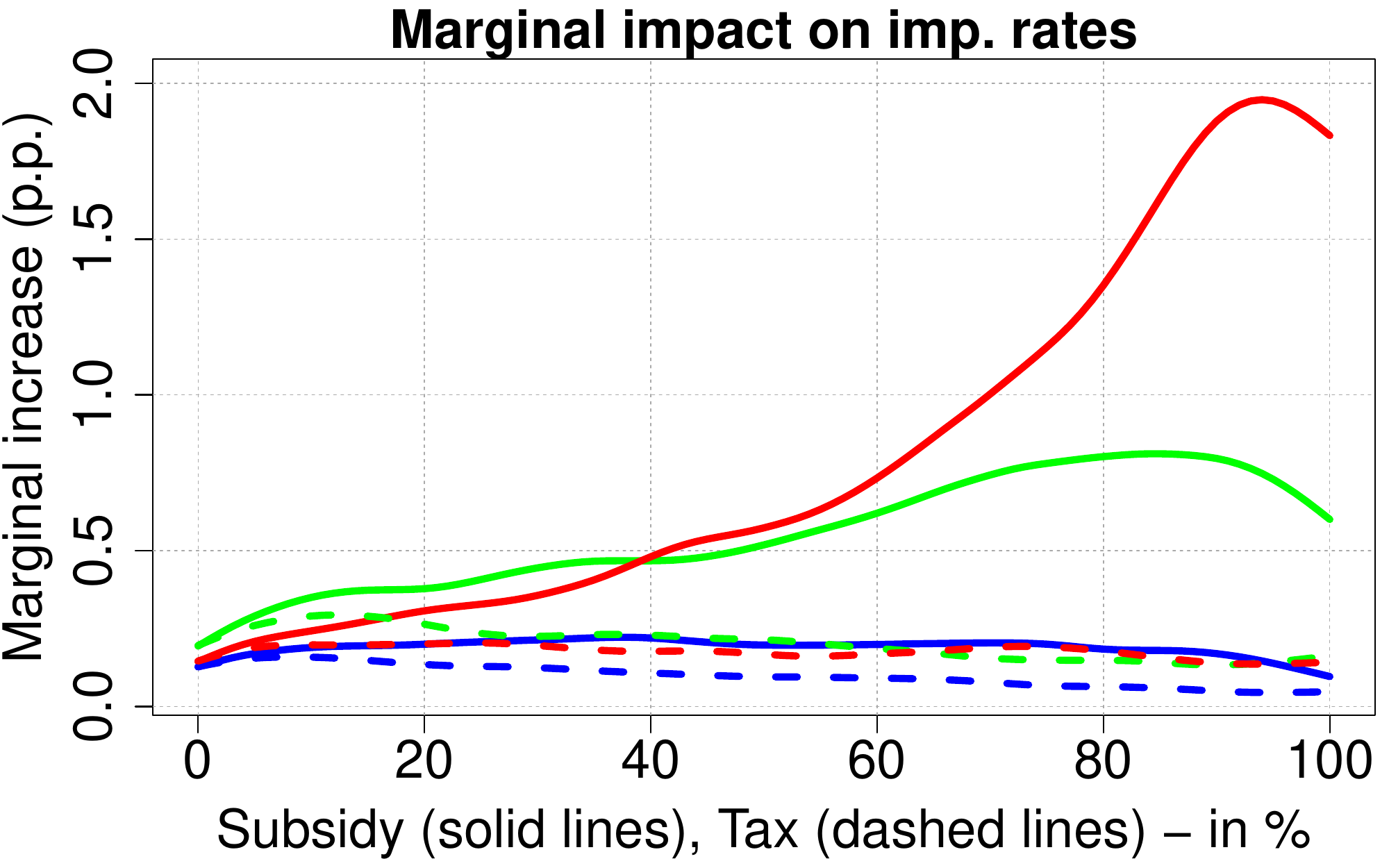}
		\end{center}
	\end{minipage}
		\begin{minipage}[t]{1\columnwidth}
		\begin{center}
		        \includegraphics[width=0.6\columnwidth]{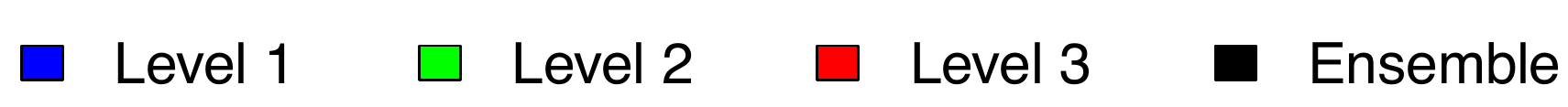}
		\end{center}
	\end{minipage}		
	\caption{Model simulations for policy effectiveness --- Implementation rates (\%), simulated NPV/NPB (thousands US\$) and marginal impact on implementation rates (p.p.). Absolute US\$ values are average values (for the tax: sum of discounted tax savings). Colours represent the different levels of decision-making of "optimizing" (blue), "satisficing" (green), "behavioural" (red) and ensemble (black, see text for details).}
	\label{fig:Figure5}
\end{figure}

\begin{figure}[t]
	\begin{center}
		\includegraphics[width=.55\columnwidth]{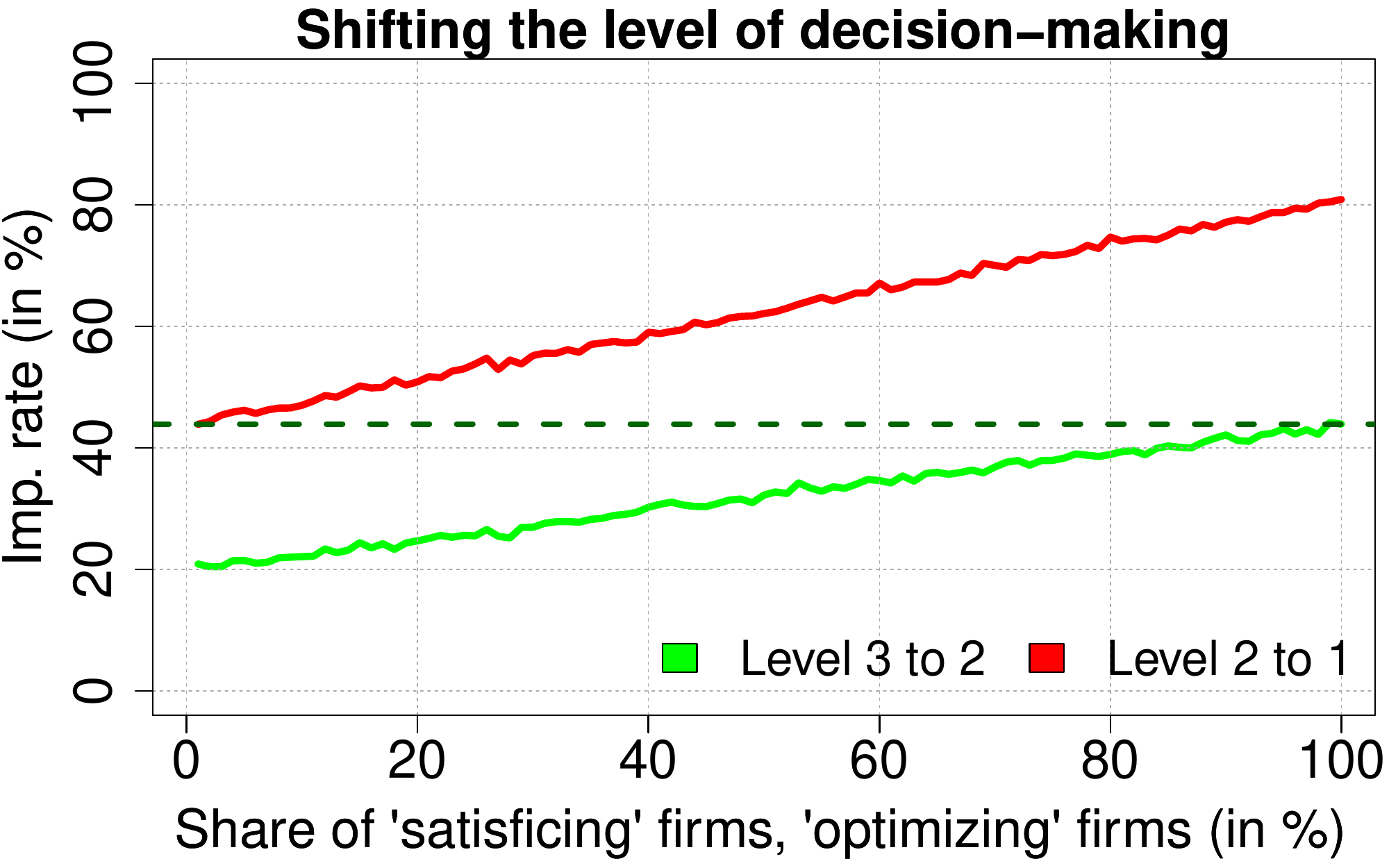}
	\end{center}
	\caption{Simulation of policies that aim at changing the level of decision-making of firms (or evaluation method): predicted implementation rate for an ensemble in which firms first gradually change from the "behavioural" level 3 to the "satisficing" level 2 (green line), and then from level 2 to the "optimizing" level 1 (red line).}
	\label{fig:Figure6}
\end{figure}

Simulation results for different policy levels are depicted in figure~\ref{fig:Figure5}, and exemplarily summarised in table~\ref{tab:policyresults}. Figure~\ref{fig:Figure6} illustrates the potential effect of shifting firms' level of decision-making. To allow an easier comparison between tax and policy, simulated tax rates (0---30\%) are chosen so that for the average firm, the resulting sums of discounted tax savings are in the same absolute range (between 0---10~000\$) as simulated subsidy payments (0---100\%). For an exemplary discussion, a tax rate (7\%) and a subsidy (27\%) are chosen so that average subsidy payments equal average discounted tax savings.

\begin{table}[t]\footnotesize
\begin{center}
		\begin{tabular*}{1\columnwidth}{@{\extracolsep{\fill}} l | c | c | c | c | c | c | c}
  & \multicolumn{2}{c |}{Impact on mean NPB} & \multicolumn{2}{c |}{Absolute impact on imp. rates} & \multicolumn{3}{c}{Relative impact on imp. rates} \\ 	
 \hline	
Level	& 7\% tax & 27\% subsidy & 7\% tax & 27\% subsidy & 7\% tax & 27\% subsidy & Relative effect subsidy/tax\\ 
\hline
\hline
 1 --- Optimizing & +2~812\$ & +2~821\$ & +1.2pp & +5.2pp & +1.5\% & +6.4\% & 4.3\\
 2 --- Satisficing	& +742\$ &+2~821\$ &+2.0pp & +9.6pp & +4.5\% & +21.8\% & 4.80\\
 3 --- Behavioural &	+204\$ &+1~830\$ &+1.5pp & +7.3pp & +7.7\% & +37.6\%& 4.87\\
 \hline
		\end{tabular*}
	\caption{Model simulations for the potential effectiveness of a 7\% tax and a 27\% subsidy on the mean NPB and the implementation rate in absolute terms (increase in percentage points) and relative terms (percentage increase). Policy rates are chosen so that average subsidy payments equal average discounted tax savings.}
	\label{tab:policyresults}
\end{center}
\end{table}

Both the tax and the subsidy are predicted to increase implementation rates. In comparison, subsidy payments are predicted to be more effective than an electricity tax on all levels, with the difference being largest for satisficing and behavioural decision-making. There are two reasons for this: first, the type of decision-making on levels 2 and 3 means that future tax savings are only considered within the individual payback thresholds. On level 3, the relative effectiveness of taxes is further lowered by the behavioural focus on upfront costs. As a result, taxes only increase the NPB to the full potential extent on level 1. Second, for the given range of data, the implementation rates show a higher (and increasing) sensitivity towards reduced upfront costs, while there is a lower (and decreasing) sensitivity towards price increases --- for all levels. This is why even on the optimizing level 1, the tax is less effective.
For the exemplary values of a 27\% subsidy and a 7\% tax (which increase the net-present value by the same absolute amount),  the subsidy (+5.2p.p.---+9.6p.p) is between four and five times more effective than the tax (+1.2p.p.---+2.0p.p) . 

With respect to levels, both policies' effectiveness is highest on levels 2 and 3: compared to level 1, lower initial implementation rates translate into larger sensitivities towards policies. As in the sensitivity analysis, there is an inverse relationship between a policy's potential effectiveness and the initial implementation rate without it (although the mean NPBs increase in a perfectly linear way). On level 1, with a high initial rate of 81\%, an ever large change in the mean NPB is needed for an any additional increase by 1 p.p.. 

\subsubsection*{The impact of information campaigns}

Given the different implementation rates on different levels, there is a potential alternative to monetary incentives: influencing the method how firms assess a green technology investment. If it is possible to \emph{shift} their decision-making towards a more conscious level (e.g. from behavioural to satisficing), this could potentially be more effective (and probably cheaper) than marginally incentivising decisions within a level. As an example, figure~\ref{fig:Figure6} illustrates a gradual \emph{shift} in decision-making from level~3 to 2, followed by a gradual \emph{shift} to level~1. 

Such policies could be the promotion of energy audits or training programs for professionals, both of which may counteract behavioural barriers \citep{Grubb:2014}. After energy audits in Australia, 80\% of firms reported that they at least perform a payback threshold analysis of recommended projects (while 30\% perform NPV calculations) \citep{Harris:2000}. Indeed, industrial energy audits are generally found to lower investment barriers, thereby increasing investments in energy efficiency in a cost-effective way \citep{Harris:2000, Anderson:2004, Schleich:2004, Trianni:2016}. In a recent review, \cite{Thollander:2015} conclude that subsidised audits are most cost-effective from a government's point of view. This evidence from the literature suggests that our match of observed implementation rates to the satisficing level may well be attributed to the energy audits themselves, and that real implementation rates could well be at the behavioural level without energy audits, with very low implementation rates.

Although from the modeller's and policy maker's perspective there remains uncertainty on how firms really make decisions, such \emph{shifting} policies are likely to be quite robust: if, for example, we assume that all firms are of the optimizing type (and no further \emph{shift} is possible), within a fully rational framework the policy could still increase the $\gamma$ value, thus counteracting the undervaluation of future benefits by providing more reliable information. Since they do not depend on cash transfers, such  policies have the additional advantage that they reduce the risk of `free riding' that comes along with subsidies.  

\section{Discussion and Conclusion}

The model introduced simulates technology adoption by heterogeneous firms in the context of behavioural heterogeneity and complexity. Simulations that were derived from it support the hypothesis that behavioural aspects account for the larger part of the energy efficiency gap, the gap between a normative benchmark and observed investments into green technologies. Despite there still being limited knowledge on what really drives firms' decisions, the inclusion of psychological and organisational findings increases the model's predictive power significantly compared to a classical optimisation approach. From a practical viewpoint, this allows for testing the robustness of policies \emph{ex ante}, to adjust them accordingly, and perhaps save governments time and money. From a theoretical perspective, this contributes to strengthen our understanding of what may enable or inhibit sustainability transitions from a microeconomic perspective.

The model was structured according to three levels of decision-making: optimizing, satisficing and behavioural, as well as an ensemble model that combines all three levels. For every level, the model simulated the adoption of a green technology from a macro perspective. Using a case study of energy-efficient electric motors, the model predicts an implementation rate as high as 81\% on the optimizing level 1. When taking into account satisficing and behavioural decision-making, the predicted rate is reduced to 44\% on level 2 and 20\% on level 3. The reported implementation rate of 45\% after energy audits is  therefore consistent with the model's prediction for the satisficing level, as well as the 52\% predicted by the ensemble model. Considering evidence from the literature, we conclude that energy audits likely influence firms' decision-making behaviour. In order to fully demonstrate this, further research is required on how audits influence the micro-level behaviour, especially a comparison of \emph{post audit} outcomes relative to counterfactual decision-making in the \emph{absence} of audits.

Behavioural aspects and heterogeneity significantly impact the effectiveness of market-based policies. Investment subsidies are predicted to have a relatively larger impact than an electricity tax --- mainly due to the larger focus on upfront costs, but also due to higher sensitivities towards reduced investment costs relative to increases in the electricity price, in the given range of data. In aggregate, a subsidy is predicted to be more than four times as effective for inducing technology uptake as a tax, even when both have the same (discounted) value in US\$. However, the largest effect could potentially be achieved by \emph{shifting} the investment decision to another level --- for example, by providing targeted audits to firms, aimed at changing \emph{the way with which decisions are taken}. In language of transitions theory and the multi-level perspective, this means changing the heuristics and routines for decision-making within the socio-technical regime.

With respect to the research design, the model is limited to a static analysis of present day technology adoption. It could, however, be coupled to dynamic simulations of technology diffusion, driven by adoption decision-making \citep[e.g. as in evolutionary models such as in ][]{Mercure2012, Mercure2014, Mercure2015}. Such a method could provide a firm basis for simulating the decisions of heterogeneous agents under bounded rationality and expectations, in order to derive resulting diffusion dynamics, and dynamics of evolving aggregate energy efficiency. In more general terms, this model structure can also help better understand what we mean with the `representative agent' in conventional economic analysis, by unpacking levels of understanding over agent characteristics. 

It is clear that when invoking an explanation for observed behaviour using the generalisation of a micro-model, outcomes may become driven by the choice of agent characteristics and/or decision rules. Therefore, the rules chosen must be the simplest and most general possible that produce the model richness required to explain existing data. It is possible that some macro-behaviour aspects are the result of other unexplored emergent properties from the complex underlying system \citep[e.g. see][]{Kirman1992}. Although here we find this unlikely, more empirical research is needed, in order to clearly demonstrate whether other decision rules could explain better our data (e.g. information asymmetry, information cascades, social influence, etc).

Policy-makers need to be made better aware that green technology investments are governed by complex behavioural processes, and that firms behave in different ways. Four key insights can be taken home for improved policy-making: (1) Behavioural aspects have a considerable effect on green technology investments. (2)  The large gap between normative-optimizing and positive analysis (the \emph{energy efficiency gap}) is at least partly predictable based on the degree of agent heterogeneity and known behavioural barriers. A model simulation based on randomised distributions can provide the policy maker with an estimate. (3) A consideration of behavioural factors can increase the robustness of policies. (4) Policies aimed at influencing the process and method of decision-making can be at least as effective, and likely more effective, than financial incentives, to increase rates of adoption of green investments.

\section*{Acknowledgements}

The authors acknowledge the German National Academic Foundation (FK) and the UK's Engineering and Physical Sciences Research Council (JFM, fellowship no EP/ K007254/1) for financial support. We gratefully thank three assiduous referees who helped us improve and clarify this manuscript significantly. We thank the Cambridge Centre for Climate Change Mitigation Research and its entire team, notably Pablo Salas. FK designed and executed the research. JFM contributed to conceptualising the research and writing the paper.


\bibliographystyle{elsarticle-harv}
\bibliography{bibliography}

\begin{thebibliography}{66}
\expandafter\ifx\csname natexlab\endcsname\relax\def\natexlab#1{#1}\fi
\expandafter\ifx\csname url\endcsname\relax
  \def\url#1{\texttt{#1}}\fi
\expandafter\ifx\csname urlprefix\endcsname\relax\def\urlprefix{URL }\fi

\bibitem[{Akerlof(1970)}]{Akerlof:1970}
Akerlof, G.~A., 1970. The market for 'lemons': Quality uncertainty and the
  market mechanism. The Quarterly Journal of Economics 84~(3), 488--500.

\bibitem[{Akerlof and Yellen(1985)}]{Akerlof:1985}
Akerlof, G.~A., Yellen, J.~L., 1985. Can small deviations from rationality make
  significant differences to economic equilibria? The American Economic Review
  75~(4), 708--720.

\bibitem[{Allais(1953)}]{Allais:1953}
Allais, M., 1953. Le comportement de l'homme rationnel devant le risque:
  Critique des postulats et axiomes de l'ecole americaine. Econometrica 21~(4),
  503--564.

\bibitem[{Allan et~al.(2014)Allan, Jaffe, and Sin}]{Allan:2014}
Allan, C., Jaffe, A.~B., Sin, I., 2014. Diffusion of green technology: A
  survey. International Review of Environmental and Resource Economics 14~(4),
  1--33.

\bibitem[{Allcott and Greenstone(2012)}]{Allcott:2012}
Allcott, H., Greenstone, M., 2012. Is there an energy efficiency gap? Working
  Paper No. w17766, National Bureau of Economic Research.

\bibitem[{Anderson and Newell(2004)}]{Anderson:2004}
Anderson, S.~T., Newell, R.~G., 2004. Information programs for technology
  adoption: the case of energy-efficiency audits. Resource and Energy Economics
  26~(1), 27--50.

\bibitem[{Arthur(1989)}]{Arthur:1989}
Arthur, W.~B., 1989. Competing technologies, increasing returns, and lock-in by
  historical events. The Economic Journal 99~(394), 116--31.

\bibitem[{Benartzi and Thaler(1995)}]{Benartzi:1995}
Benartzi, S., Thaler, R., 1995. Myopic loss aversion and the equity premium
  puzzle. Quarterly Journal of Economics 110~(1), 73--92.

\bibitem[{{C}enter for {A}dvanced~{E}nergy {S}tudies(2015)}]{CAES:2015}
{C}enter for {A}dvanced~{E}nergy {S}tudies, 2015. Industrial assessment centers
  database. Database, {R}utgers {U}niversity.
\newline\urlprefix\url{http://iac.rutgers.edu}

\bibitem[{Crawford-Brown(1999)}]{Crawford-Brown:1999}
Crawford-Brown, D., 1999. Risk-Based Environmental Decisions. Kluwer Academic
  Publishers.

\bibitem[{Damodaran(2007)}]{Damodaran:2007}
Damodaran, A., 2007. Strategic Risk Taking: A Framework for Risk Management.
  Pearson Prentice Hall.

\bibitem[{DeCanio(1998)}]{DeCanio:1998}
DeCanio, S.~J., 1998. The efficiency paradox: Bureaucratic and organizational
  barriers to profitable energy-saving investments. Energy Policy 26~(5),
  441--454.

\bibitem[{Ellsberg(1961)}]{Ellsberg:1961}
Ellsberg, D., 1961. Risk, ambiguity, and the savage axioms. The Quarterly
  Journal of Economics 75~(4), 643--669.

\bibitem[{Fisher and Pry({1971})}]{Fisher1971}
Fisher, J.~C., Pry, R.~H., {1971}. {A simple substitution model of
  technological change}. {Technological Forecasting and Social Change}
  {3}~({1}), {75--88}.

\bibitem[{Geels(2002)}]{Geels:2002}
Geels, F.~W., 2002. Technological transitions as evolutionary reconfiguration
  processes: a multi-level perspective and a case-study. Research Policy
  31~(8), 1257--1274.

\bibitem[{Geels(2005)}]{Geels2005}
Geels, F.~W., 2005. The dynamics of transitions in socio-technical systems: A
  multi-level analysis of the transition pathway from horse-drawn carriages to
  automobiles (1860 - 1930). Technology Analysis \& Strategic Management
  17~(4), 445--476.

\bibitem[{Geels(2011)}]{Geels2011}
Geels, F.~W., 2011. The multi-level perspective on sustainability transitions:
  Responses to seven criticisms. Environmental innovation and societal
  transitions 1~(1), 24--40.

\bibitem[{Genus and Coles(2008)}]{Genus2008}
Genus, A., Coles, A.-M., 2008. Rethinking the multi-level perspective of
  technological transitions. Research policy 37~(9), 1436--1445.

\bibitem[{Gilchrist et~al.(2013)Gilchrist, Sim, and Zakraj{\v
  s}ek}]{Gilchrist:2013}
Gilchrist, S., Sim, J.~W., Zakraj{\v s}ek, E., 2013. Misallocation and
  financial market frictions: Some direct evidence from the dispersion in
  borrowing costs. Review of Economic Dynamics 16~(1), 159--176.

\bibitem[{Gillingham et~al.(2009)Gillingham, Newell, and
  Palmer}]{Gillingham:2009}
Gillingham, K., Newell, R.~G., Palmer, K., 2009. Energy efficiency economics
  and policy. Working Paper No. w15031, National Bureau of Economic Research.

\bibitem[{Gillingham and Palmer(2014)}]{Gillingham:2014}
Gillingham, K., Palmer, K., 2014. Bridging the energy efficiency gap: Policy
  insights from economic theory and empirical evidence. Review of Environmental
  Economics and Policy 8~(1), 18--38.

\bibitem[{Graham and Harvey(2001)}]{Graham:2001}
Graham, J.~R., Harvey, C.~R., 2001. The theory and practice of corporate
  finance: Evidence from the field. Journal of Financial Economics 60~(2),
  187--243.

\bibitem[{Greene(2011)}]{Greene:2011}
Greene, D.~L., 2011. Uncertainty, loss aversion, and markets for energy
  efficiency. Energy Economics 33~(4), 608--616.

\bibitem[{Grubb(2014)}]{Grubb:2014}
Grubb, M., 2014. Planetary economics: energy, climate change and the three
  domains of sustainable development. Taylor Francis/Routledge.

\bibitem[{Gr{\"u}bler and Naki{\'c}enovi{\'c}(1991)}]{Grubler:1991}
Gr{\"u}bler, A., Naki{\'c}enovi{\'c}, N., 1991. Long waves, technology
  diffusion, and substitution. Review XIV~(2), 313--342.

\bibitem[{Gr\"ubler et~al.(1999)Gr\"ubler, Nakicenovic, and
  Victor}]{Grubler1999}
Gr\"ubler, A., Nakicenovic, N., Victor, D., 1999. Dynamics of energy
  technologies and global change. Energy Policy 27~(5), 247--280.

\bibitem[{Harris et~al.(2000)Harris, Anderson, and Shafron}]{Harris:2000}
Harris, J., Anderson, J., Shafron, W., 2000. Investment in energy efficiency: a
  survey of australian firms. Energy Policy 28~(12), 867--876.

\bibitem[{Hassett and Metcalf(1995)}]{Hassett:1995}
Hassett, K.~A., Metcalf, G.~E., 1995. Energy tax credits and residential
  conservation investment: Evidence from panel data. Journal of Public
  Economics 57~(2), 201--217.

\bibitem[{Hofbauer and Sigmund(1998)}]{Hofbauer1998}
Hofbauer, J., Sigmund, K., 1998. Evolutionary Games and Population Dynamics.
  Cambridge University Press, Cambridge, UK.

\bibitem[{Holtz(2011)}]{Holtz2011}
Holtz, G., 2011. Modelling transitions: An appraisal of experiences and
  suggestions for research. Environmental Innovation and Societal Transitions
  1~(2), 167 -- 186.

\bibitem[{Holtz et~al.(2015)Holtz, Alkemade, de~Haan, K√∂hler, Trutnevyte,
  Luthe, Halbe, Papachristos, Chappin, Kwakkel, and Ruutu}]{Holtz2015}
Holtz, G., Alkemade, F., de~Haan, F., K√∂hler, J., Trutnevyte, E., Luthe, T.,
  Halbe, J., Papachristos, G., Chappin, E., Kwakkel, J., Ruutu, S., 2015.
  Prospects of modelling societal transitions: Position paper of an emerging
  community. Environmental Innovation and Societal Transitions, in press, --.

\bibitem[{Jaffe et~al.(2005)Jaffe, Newell, and Stavins}]{Jaffe:2005}
Jaffe, A.~B., Newell, R.~G., Stavins, R.~N., 2005. A tale of two market
  failures: Technology and environmental policy. Ecological Economics 54~(2),
  164--174.

\bibitem[{Jaffe and Stavins(1994)}]{Jaffe:1994}
Jaffe, A.~B., Stavins, R.~N., 1994. The energy-efficiency gap. what does it
  mean? Energy Policy 22~(10), 804--810.

\bibitem[{Kahneman(2003)}]{Kahneman:2003}
Kahneman, D., 2003. Maps of bounded rationality: Psychology for behavioral
  economics. American Economic Review 93~(5), 1449--1475.

\bibitem[{Kahneman and Tversky(1979)}]{Kahneman:1979}
Kahneman, D., Tversky, A., 1979. Prospect theory: An analysis of decision under
  risk. Econometrica 47~(2), 263--292.

\bibitem[{Kirman(1992)}]{Kirman1992}
Kirman, A.~P., 1992. Whom or what does the representative individual represent?
  The Journal of Economic Perspectives, 117--136.

\bibitem[{Mansfield(1961)}]{Mansfield1961}
Mansfield, E., 1961. Technical change and the rate of imitation. Econometrica
  29~(4), pp. 741--766.

\bibitem[{Marchetti and Nakicenovic(1978)}]{Marchetti1978}
Marchetti, C., Nakicenovic, N., 1978. The dynamics of energy systems and the
  logistic substitution model. Tech. rep., IIASA.
\newline\urlprefix\url{http://www.iiasa.ac.at/Research/TNT/WEB/PUB/RR/rr-79-13.pdf}

\bibitem[{{McKinsey and Company}(2009)}]{McKinsey:2009}
{McKinsey and Company}, 2009. Pathways to a low-carbon economy: Version 2 of
  the global greenhouse gas abatement cost curve.
\newline\urlprefix\url{http://www.mckinsey.com/~/media/mckinsey/dotcom/client_service/sustainability/cost\%20curve\%20pdfs/pathways_lowcarbon_economy_version2.ashx}

\bibitem[{Mercure(2012)}]{Mercure2012}
Mercure, J.-F., 2012. Ftt:power : A global model of the power sector with
  induced technological change and natural resource depletion. Energy Policy
  48~(0), 799 -- 811.
\newline\urlprefix\url{http://dx.doi.org/10.1016/j.enpol.2012.06.025}

\bibitem[{Mercure(2015)}]{Mercure2015}
Mercure, J.-F., 2015. An age structured demographic theory of technological
  change. J. Evol. Econ. 25, 787--820.

\bibitem[{Mercure et~al.(2014)Mercure, Pollitt, Chewpreecha, Salas, Foley,
  Holden, and Edwards}]{Mercure2014}
Mercure, J.-F., Pollitt, H., Chewpreecha, U., Salas, P., Foley, A., Holden, P.,
  Edwards, N., 2014. The dynamics of technology diffusion and the impacts of
  climate policy instruments in the decarbonisation of the global electricity
  sector. Energy Policy 73~(0), 686 -- 700.

\bibitem[{Metcalfe(1988)}]{Metcalfe1988}
Metcalfe, J.~S., 1988. {The diffusion of innovations: an interpretative
  survey}. In: Dosi, G., Freeman, C., Nelson, R., Silverberg, G., Soete, L.
  (Eds.), Technical change and economic theory. Pinter Publishers, pp.
  560--607.

\bibitem[{Nakicenovic({1986})}]{Nakicenovic1986}
Nakicenovic, N., {1986}. {The automobile road to technological-change -
  Diffusion of the automobile as a process of technological substitution}.
  {Technological Forecasting and Social Change} {29}~({4}), {309--340}.

\bibitem[{Pollitt and Shaorshadze(2012)}]{Pollitt:2012}
Pollitt, M., Shaorshadze, I., 2012. The role of behavioural economics in energy
  and climate policy. In: Fouquet, R. (Ed.), Handbook of energy and climate
  change. Edward Elgar Publications.

\bibitem[{Rogers(2010)}]{Rogers:2010}
Rogers, E.~M., 2010. Diffusion of Innovations. Simon and Schuster.

\bibitem[{Rotmans et~al.(2001)Rotmans, Kemp, and Van~Asselt}]{Rotmans2001}
Rotmans, J., Kemp, R., Van~Asselt, M., 2001. More evolution than revolution:
  Transition management in public policy. Foresight 3~(1), 15--31.

\bibitem[{Safarzy{\'n}ska et~al.(2012)Safarzy{\'n}ska, Frenken, and van~den
  Bergh}]{Safarzynska2012b}
Safarzy{\'n}ska, K., Frenken, K., van~den Bergh, J. C. J.~M., 2012.
  Evolutionary theorizing and modeling of sustainability transitions. Research
  Policy 41~(6), 1011--1024.

\bibitem[{Safarzynska and van~den Bergh({2010})}]{Safarzynska2010}
Safarzynska, K., van~den Bergh, J. C. J.~M., {2010}. {Evolutionary models in
  economics: a survey of methods and building blocks}. Journal of Evolutionary
  Economics {20}~({3}), {329--373}.

\bibitem[{Safarzynska and van~den Bergh(2012)}]{Safarzynska2012}
Safarzynska, K., van~den Bergh, J. C. J.~M., 2012. An evolutionary model of
  energy transitions with interactive innovation-selection dynamics. Journal of
  Evolutionary Economics, 1--23.

\bibitem[{Saviotti and Mani(1995)}]{Saviotti1995}
Saviotti, P., Mani, G., 1995. Competition, variety and technological evolution:
  A replicator dynamics model. Journal of Evolutionary Economics 5~(4),
  369--392.

\bibitem[{Saviotti(1991)}]{Saviotti1991}
Saviotti, P.~P., 1991. {The role of variety in economic and technological
  development}. In: Saviotti, P.~P., Metcalfe, J.~S. (Eds.), Evolutionary
  theories of economic and technological change. Harwood Academic Publishers,
  New York, USA, pp. 133--159.

\bibitem[{Schleich(2004)}]{Schleich:2004}
Schleich, J., 2004. Do energy audits help reduce barriers to energy efficiency?
  an empirical analysis for germany. International Journal of Energy Technology
  2~(3), 226--239.

\bibitem[{Shogren and Taylor(2008)}]{Shogren:2008}
Shogren, J.~F., Taylor, L.~O., 2008. On behavioral-environmental economics.
  Review of Environmental Economics and Policy 2~(1), 26--44.

\bibitem[{Simon(1955)}]{Simon:1955}
Simon, H.~A., 1955. A behavioral model of rational choice. The Quarterly
  Journal of Economics 69~(1), 99--118.

\bibitem[{Sorrell(2004)}]{Sorrell:2004}
Sorrell, S., 2004. The Economics of Energy Efficiency: Barriers to
  Cost-Effective Investment. Edward Elgar Publishing.

\bibitem[{Sorrell et~al.(2011)Sorrell, Mallett, and Nye}]{Sorrell:2011}
Sorrell, S., Mallett, A., Nye, S., 2011. Barriers to industrial energy
  efficiency: A literature review. Working Paper 10/2011, United Nations
  Industrial Development Organization.
\newline\urlprefix\url{https://www.unido.org/fileadmin/user_media/Services/Research_and_Statistics/WP102011_Ebook.pdf}

\bibitem[{Sutherland(1991)}]{Sutherland:1991}
Sutherland, R.~J., 1991. Market barriers to energy-efficiency investments. The
  Energy Journal 12~(3), 15--34.

\bibitem[{Thollander et~al.(2015)Thollander, Kimura, Wakabayashi, and
  Rohdin}]{Thollander:2015}
Thollander, P., Kimura, O., Wakabayashi, M., Rohdin, P., 2015. A review of
  industrial energy and climate policies in japan and sweden with emphasis
  towards smes. Renewable and Sustainable Energy Reviews 50, 504--512.

\bibitem[{Trianni et~al.(2016)Trianni, Cagno, and Farne}]{Trianni:2016}
Trianni, A., Cagno, E., Farne, S., 2016. Barriers, drivers and decision-making
  process for industrial energy efficiency: A broad study among manufacturing
  small and medium-sized enterprises. Applied Energy 162, 1537–1551.

\bibitem[{Tversky and Kahneman(1992)}]{Tversky:1992}
Tversky, A., Kahneman, D., 1992. Advances in prospect theory: Cumulative
  representation of uncertainty. Journal of Risk and Uncertainty 5~(4),
  297--323.

\bibitem[{von Neumann and Morgenstern(1947)}]{Von-Neumann:1947}
von Neumann, J., Morgenstern, O., 1947. The Theory of Games and Economic
  Behavior. Princeton University Press.

\bibitem[{Waide and Brunner(2011)}]{IEA:2011}
Waide, P., Brunner, C., 2011. Energy-efficiency policy opportunities for
  electric motor-driven systems. Energy efficiency series working paper,
  International Energy Agency.
\newline\urlprefix\url{http://www.oecd-ilibrary.org/content/workingpaper/5kgg52gb9gjd-en}

\bibitem[{Williamson(1967)}]{Williamson:1967}
Williamson, O.~E., 1967. Hierarchical control and optimum firm size. The
  Journal of Political Economy 75~(2), 123--138.

\bibitem[{Wilson and Grubler(2011)}]{Wilson:2011}
Wilson, C., Grubler, A., 2011. Lessons from the history of technological change
  for clean energy scenarios and policies. Natural Resources Forum 35~(3),
  165--184.

\bibitem[{Worrell et~al.(2009)Worrell, Bernstein, Roy, Price, and
  Harnisch}]{Worrell:2009}
Worrell, E., Bernstein, L., Roy, J., Price, L., Harnisch, J., 2009. Industrial
  energy efficiency and climate change mitigation. Energy Efficiency 2~(2),
  109--123.

\end{thebibliography}

\end{document}